\documentclass[acmsmall,screen,nonacm]{acmart}

\usepackage{xspace}
\usepackage{array}
\usepackage{multirow}
\usepackage{algorithm}
\usepackage{algpseudocode}
\usepackage{xcolor}
\usepackage{colortbl}
\usepackage{placeins}

\definecolor{tablecanonical}{HTML}{ECEFF1}
\definecolor{tablecg}{HTML}{F2F3F2}
\definecolor{tabletfg}{HTML}{E4ECF4}
\definecolor{tableunion}{HTML}{F3EBDC}
\definecolor{tableheader}{HTML}{ECEFF1}
\definecolor{tablesection}{HTML}{F5F6F7}
\definecolor{tablehighlight}{HTML}{EAF1F7}
\definecolor{tablesubtotal}{HTML}{F1F5F8}
\definecolor{tablegrand}{HTML}{DCE4EC}

\AtBeginDocument{%
  }

\setcopyright{none}
\newcommand{\tool}{FlowMuT\xspace}
\newcommand{\tfg}{TFG\xspace}

\newcommand{\parlabel}[1]{%
  \noindent\hspace*{\parindent}\textbf{#1}}
\newcommand{\rqanswer}[2]{%
  \par\addvspace{\smallskipamount}
  \begingroup
  \setlength{\fboxsep}{4pt}%
  \setlength{\fboxrule}{0.4pt}%
  \noindent\fcolorbox{black!55}{black!3}{%
    \begin{minipage}{\dimexpr\linewidth-2\fboxsep-2\fboxrule\relax}
      \textbf{Answer to #1.} #2
    \end{minipage}}%
  \endgroup
  \par\addvspace{\smallskipamount}
}
\newcolumntype{P}[1]{>{\raggedright\arraybackslash}p{#1}}
\newcolumntype{C}[1]{>{\centering\arraybackslash}p{#1}}
\newcolumntype{R}[1]{>{\raggedleft\arraybackslash}p{#1}}

\begin{document}

\title{Grounding Deep Learning Framework Testing in Data Flow}

\author{Jiatao Ge}
\orcid{0009-0009-9394-8100}
\affiliation{%
  \institution{State Key Laboratory for Novel Software Technology, Nanjing University}
  \city{Nanjing}
  \country{China}}
\email{gejiatao@smail.nju.edu.cn}

\author{Yanzhou Mu}
\orcid{0000-0003-1816-2246}
\affiliation{%
  \institution{UNIST}
  \city{Ulsan}
  \country{Republic of Korea}}
\email{602022320006@smail.nju.edu.cn}

\author{Yinglong Zou}
\orcid{0009-0006-9375-7417}
\affiliation{%
  \institution{State Key Laboratory for Novel Software Technology, Nanjing University}
  \city{Nanjing}
  \country{China}}
\email{652023320004@smail.nju.edu.cn}

\author{Yibo Hu}
\orcid{0009-0009-4103-1421}
\affiliation{%
  \institution{State Key Laboratory for Novel Software Technology, Nanjing University}
  \city{Nanjing}
  \country{China}}
\email{yibohu@smail.nju.edu.cn}

\author{Zhenyu Chen}
\orcid{0000-0002-9592-7022}
\affiliation{%
  \institution{State Key Laboratory for Novel Software Technology, Nanjing University}
  \city{Nanjing}
  \country{China}}
\email{zychen@nju.edu.cn}

\author{Chunrong Fang}
\orcid{0000-0002-9930-7111}
\authornote{Corresponding author.}
\affiliation{%
  \institution{State Key Laboratory for Novel Software Technology, Nanjing University}
  \city{Nanjing}
  \country{China}}
\email{fangchunrong@nju.edu.cn}
\renewcommand{\shortauthors}{Ge et al.}

\begin{abstract}
Data flow is important to many aspects of deep learning (DL) framework
execution, such as automatic differentiation and operator-fusion
optimization.
Framework bugs involving data flow may therefore cause incorrect results or
other unexpected behavior.  Detecting such bugs, however, is challenging for
existing testing methods that represent models primarily by operator topology.
This representation records operators and their connections but omits runtime
tensor values and properties as well as operator input requirements.  Without
this information, these methods cannot distinguish different data flows in
models with similar topologies, even when those differences affect framework
behavior.
To address this gap, we introduce the Tensor Flow Graph (\tfg) to capture the
runtime tensor states of model test cases and their operators' input
requirements.  Its nodes represent operators, and its edges represent tensors
flowing between them.  Each edge records the runtime values and
properties of its tensor, while each node records the input requirements of its
operator.  Building on the \tfg, we propose \tool, a mutation-based testing
method grounded in data flow.  We represent mutations as constrained subgraph
replacements in the TFG.  \tool mutation operators are derived from prior work
and an analysis of mutable data flow objects.
\tool iteratively mutates a seed model and tests each resulting mutant.  In
each round, it builds a TFG for the current seed model, pairs mutation operators
with TFG sites to form candidates, and filters illegal candidates using data
flow constraints.  It then uses LinUCB
to select a candidate based on coverage and impact.  After applying the
selected mutation and executing the mutant, it updates selection feedback and
uses each accepted mutant as the seed model for the next round.
We evaluate \tool on PyTorch and MindSpore using representative models from
diverse application scenarios.  \tool submitted 55 bug reports for cases
missed by all baseline methods.  Developers confirmed 41 reports and fixed
the bugs described in 7 reports.
\tool generates more diverse test cases, achieving 43.4\% higher data flow
diversity than the strongest baseline.  It achieves a legal mutant rate of
96.7\%, compared with 86.4\% without constraints.  These results show that testing
from a data flow perspective is practical and effective.
\end{abstract}

\ccsdesc[500]{Software and its engineering~Software testing and debugging}

\keywords{deep learning framework testing, mutation-based testing, data flow}

\maketitle

\begin{figure}[!t]
  \centering
  \includegraphics[width=\linewidth]{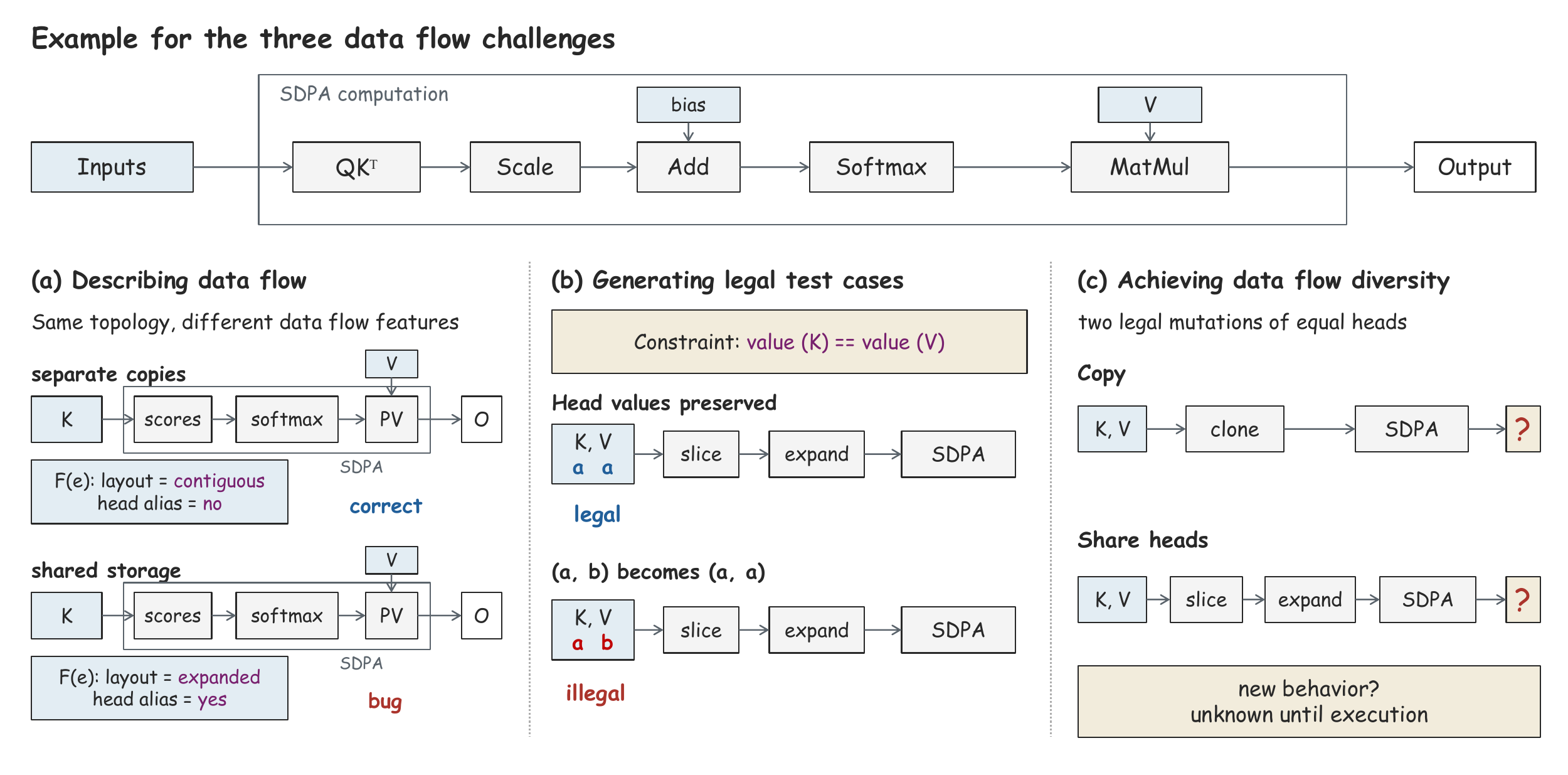}
  \caption{Three data flow challenges illustrated by scaled dot-product attention (SDPA).}
  \Description{The top panel shows a scaled dot-product attention computation.
  Panel (a) compares a correct execution with separate contiguous copies of
  K and V against a buggy execution with expanded, aliased storage under the
  same topology.  Panel (b) shows that slice-and-expand head sharing preserves
  values for equal K and V heads but changes unequal heads, making that
  candidate illegal.  Panel (c) shows two legal mutations of equal heads,
  cloning and sharing, whose behavioral novelty is unknown.}
  \label{fig:running-example}
\end{figure}

\section{Introduction}

Data flow encompasses the tensors passed between operators and their runtime
values and properties in deep learning (DL) framework execution.  Automatic
differentiation depends on data flow because it traverses the executed graph
in reverse and reuses intermediate tensor values to propagate
gradients~\cite{abadi2016tensorflow,paszke2019pytorch}.  Operator fusion also
depends on data flow because it analyzes the executed graph and its tensor
properties to determine which operator groups can be
fused~\cite{chen2018tvm,rotem2018glow}.  Framework bugs involving data flow
may disrupt these and other mechanisms.  Incorrect intermediate values can
affect gradient propagation, while an invalid fusion may change the
computation represented by the original graph.  Such bugs may cause incorrect
results or erroneous behavior.  Therefore, detecting them is
important for DL framework testing~\cite{mu2026llmchallenges}.

\begingroup\looseness=-1
Existing DL framework testing methods construct and execute test cases to
detect framework bugs.  Here, a test case is a DL model.  Such a model is
commonly represented as a computational graph (CG)
that records its operator topology, which consists of its operators and their
connections~\cite{wang2020lemon,guo2020audee,gu2022muffin,li2023comet,
liu2023nnsmith,mu2024devmut}.  This description does not capture
runtime tensor values and properties or operator input requirements, leaving
test cases unorganized around data flow.\par\endgroup

Grounding DL framework testing in data flow presents three connected
challenges.  Figure~\ref{fig:running-example} uses a scaled dot-product
attention (SDPA) computation to illustrate these challenges.

\parlabel{Challenge 1: Describing data flow.} Existing methods describe a model
using its CG, which records the operators and their connections but not the tensor
values and properties (such as shapes, layouts, alias relations, and gradient
states) that determine data flow behavior.  Two model executions with the same
CG may exercise very different data flow behavior.  Without such data flow
information, framework behavior under particular data flow states may remain
untested.  As Figure~\ref{fig:running-example}(a) shows, two executions of the
same SDPA topology differ in tensor layout and storage aliasing.  With separate
contiguous copies, the computation produces the expected result; with expanded
tensors that share storage, it exposes a framework bug.  A CG records the same
operators and connections in both cases and cannot capture the data flow state
associated with the failure.

\parlabel{Challenge 2: Generating legal test cases.} Existing methods generate
new test cases by applying mutation operators to the
structures of seed models~\cite{wang2020lemon,guo2020audee,
gu2022muffin,li2023comet,mu2024devmut}.  However, they overlook the complexity
of the data flow surrounding mutation sites, where multiple tensors are
produced and consumed subject to many interdependent constraints.  Among other
aspects, these constraints govern dtype and shape compatibility, operator
input requirements, and whether a mutation preserves the intended tensor
values; they must be satisfied simultaneously.  Consequently, they generate a
large proportion of illegal test cases whose failures do not indicate
framework bugs.  In Figure~\ref{fig:running-example}(b), a slice-and-expand
mutation shares the $K$ and $V$ heads.  To preserve their values, this mutation
requires the two heads to be equal.  It is legal for $(a,a)$, but changes
$(a,b)$ to $(a,a)$ when the values differ and must therefore be rejected.

\parlabel{Challenge 3: Achieving data flow diversity.} A seed model
can be mutated in many ways, but only a small subset of these mutations can
exercise such behavior.  Existing methods guide mutation selection using
structural information about the model, but do not consider whether a mutation will
exercise new data flow behavior.  As a result, multiple test cases may
repeatedly exercise similar data flow behavior, leaving other behavior
untested.  As Figure~\ref{fig:running-example}(c) shows, two mutations of
equal heads satisfy the constraints.  One clones them into separate storage,
while the other shares them using slice and expand.  Both preserve the original
values but produce different layout and aliasing states.  Whether either
mutation exposes previously unseen framework behavior is unknown.

To address these challenges, we introduce the Tensor Flow Graph (\tfg), which
represents a profiled model execution as operator nodes connected by tensor
edges.  Each edge records its tensor's runtime values
and properties, and each node records its operator's input requirements.  On
the \tfg,
mutations are represented uniformly as constrained subgraph replacements
\textbf{(addressing Challenge 1)}.  Building on the \tfg, we propose \tool, a
novel testing method grounded in data flow.  Its operator pool combines
CG-derived operators adapted from prior
work with TFG-derived operators constructed from TFG objects and mutation
primitives, yielding 147 mutation operators.  In each mutation round, \tool
first builds the TFG of the current seed model, constructs mutation
candidates by pairing operators with TFG sites using a novel and efficient
algorithm, and
filters illegal candidates using input, internal, and output constraints
\textbf{(addressing Challenge 2)}.  It then treats candidate selection as a
bandit problem and uses LinUCB, a reinforcement learning method, to select a
candidate based on coverage and impact \textbf{(addressing Challenge 3)}.
Finally, it applies and executes the selected candidate, checks for framework
failures, updates the selection feedback, and uses each accepted mutant as the
seed model for the next round.

\begingroup\looseness=-1
We evaluate \tool on PyTorch and MindSpore using 14 representative models from
ten application scenarios.  We compare it with three state-of-the-art DL
framework testing methods
(LEMON~\cite{wang2020lemon}, COMET~\cite{li2023comet}, and
DevMuT~\cite{mu2024devmut}).  The results show that \tool detects 79 bugs,
compared with 28 for the strongest baseline.  We submitted 55 bug reports for
cases detected only by \tool.  Framework developers confirmed 41 reports and
fixed the bugs described in 7 reports.
Analysis of a MindSpore crash detected by \tool also revealed two
denial-of-service (DoS) vulnerabilities, both acknowledged by MindSpore's
security team.
\tool also achieves the highest data flow diversity among the evaluated
methods, exceeding COMET, the strongest baseline on this metric, by 43.4\%.
It achieves a legal mutant rate of 96.7\% with data flow constraints, compared
with 86.4\% without constraints.  These results show that testing from a data
flow perspective is practical and effective.\par\endgroup

In summary, the main contributions of this work are as follows.

\begingroup
\setlength{\listisep}{0pt}
\begin{itemize}
  \item \textbf{Data Flow Representation.} We introduce the \tfg to organize DL
  framework testing around operator topology, runtime tensor states, and
  operator input requirements.
  \item \textbf{Mutation-Based Testing Method.} We propose \tool, which
  constructs and filters candidates from CG-derived and TFG-derived operators
  and uses LinUCB with execution feedback to guide selection.
  \item \textbf{Empirical Evaluation.} We show that \tool detects framework bugs
  missed by prior methods while improving data flow diversity and mutation
  legality.
  \item \textbf{Open Research Artifact.} We release the implementation,
  mutation operators, experimental materials, mutants, triggering test cases,
  and bug reports for reproduction~\cite{flowmut2026artifact}.
\end{itemize}
\endgroup

\section{Background and Related Work}
\label{sec:related}

\subsection{Data Flow in DL Frameworks}

DL frameworks use data flow to execute models, propagate gradients, and
optimize computations.
Executing a DL model produces a sequence of tensors that flow between
operators.  The \emph{data flow} of an execution consists of these produced
and consumed tensors together with their runtime values and properties, such
as shape, layout, numerical state, and gradient state.  We distinguish this
dynamic data flow behavior from \emph{operator topology}, which records the
operators in a model and their static connections.  Operator
topology describes the routes along which tensors may travel, whereas data
flow describes what actually travels along those routes in a particular
execution.  Consequently, the same operator topology can realize different
data flows under different inputs and execution conditions.  The following
two examples illustrate the role of data flow in automatic differentiation
and in compilation and optimization.

\parlabel{Automatic differentiation.}
Automatic differentiation follows the data flow produced during forward
execution to propagate gradients.  In PyTorch, operators construct an
execution-specific graph as they run~\cite{paszke2019pytorch}.  The
automatic-differentiation engine traverses this graph in reverse and uses
saved intermediate tensors and gradient states to compute derivatives.  Work
on testing automatic differentiation further shows that framework behavior
can differ across forward, first-order-gradient, and higher-order-gradient
execution scenarios~\cite{yang2023nablafuzz}.  Thus, gradient propagation
depends on both operator dependencies and the runtime states of their tensors.

\parlabel{Compilation and optimization.}
DL compiler research uses data flow information, including operator
dependencies and tensor properties, to transform model execution.  TVM
combines graph-level transformations, including operator fusion, with
operator-level tensor optimization~\cite{chen2018tvm}.  For fusion,
producer--consumer relations show which operators exchange tensors, while
shape and layout information inform the fused implementation.  nGraph and
Glow introduce intermediate representations that progressively lower
neural-network graphs toward executable code, enabling graph rewriting,
kernel selection, and memory optimization~\cite{cyphers2018ngraph,
rotem2018glow}.  MLIR generalizes this organization through reusable
multi-level intermediate representations, while PyTorch 2 captures executed
program regions for graph transformation and compilation~\cite{lattner2021mlir,
ansel2024pytorch2}.  These systems use operator dependencies and tensor
properties to improve execution, rather than to organize the generation of
framework test cases.

\begingroup\looseness=-1
\parlabel{Comparison with prior work.}
Prior work establishes data flow as a basis for automatic differentiation and
compiler optimization.  The corresponding graphs and
intermediate representations may record tensor types, dependencies, or other
information required by their particular execution and optimization tasks.
They are not, however, designed as a common mutation-testing abstraction.
Existing testing work also uses operator constraints and execution traces for
particular generation or checking tasks, as discussed next, but does not bring
them together with profiled runtime tensor states to organize mutation
construction, legality checking, and candidate selection.  The \tfg fills this
role by jointly recording operator topology, runtime tensor values and
properties, and operator input requirements for a profiled whole-model
execution.\par\endgroup

\subsection{DL Framework Testing}

\begingroup\looseness=-1
DL framework testing constructs test inputs, executes them, and applies an
oracle to identify crashes, incorrect results, or inconsistent
behavior~\cite{pham2019cradle,mu2026cipihunter,chen2018metamorphic,
mu2025modelmeta,zhang2022mltesting,sun2020transrepair,zhou2020metamorphic}.
Methods can be grouped by the granularity of their test inputs: API-level
testing exercises individual framework calls or short call sequences, while
model-level testing executes complete models with connected
operators~\cite{zhang2025dllibrarysurvey}.  The former probes operators under
different arguments and execution contexts; the latter also exercises their
interactions during model execution.  Existing models can serve directly as
tests, as in CRADLE's cross-backend comparison~\cite{pham2019cradle}.  To
construct new model-level tests, generation-based methods synthesize models
from templates, grammars, or operator constraints, while mutation-based
methods transform executable seed models.\par\endgroup

\parlabel{API-level testing.}
API-level methods vary tensor arguments, operator parameters, and execution
contexts, sometimes in short call sequences.  FreeFuzz mines argument types
and values from documentation, developer tests, and models to generate API
tests~\cite{wei2022freefuzz}.  Other methods generate or refine API tests
under different inputs and execution contexts~\cite{harzevili2025orion,
deng2023titanfuzz,kim2025pathfinder,naziri2026centaur,
christou2023ivysyn,xie2022docter,shi2025multigranularity}.  These tests
exercise framework APIs without requiring a complete DL model as a test case.

\parlabel{Generation-based model testing.}
Generation-based methods synthesize models from graph templates, grammars, or
operator constraints.  Muffin builds models from directed acyclic graph
structures and applies differential testing
to forward propagation, loss computation, and backward
propagation~\cite{gu2022muffin}.  NNSmith incrementally constructs valid models
using manually specified operator constraints and tensor-type propagation, and
uses gradient-guided search to avoid exceptional floating-point
values~\cite{liu2023nnsmith}.  Gandalf combines grammar-based model generation
with reinforcement learning and metamorphic relations to diversify generated
models and their executions~\cite{liu2024gandalf}.  NeuRI further reduces the
manual specification effort by inferring operator constraints from valid and
invalid API traces and combining symbolic and concrete operators during model
generation~\cite{liu2023neuri}.  Other methods specialize model or program
generation for compiler transformations, optimization passes, and particular
execution environments~\cite{liu2022tzer,su2023torchprobe,ma2023hirgen,
zhang2025orthrus,zou2024dljsfuzzer,zou2026scalpel}.

\parlabel{Mutation-based model testing.}
Building on model-level differential testing, LEMON mutates model structures
and weights and uses a heuristic strategy to favor mutants that amplify
cross-library output inconsistencies~\cite{wang2020lemon}.  Audee combines
mutations of model inputs, parameters, and weights with evolutionary search to
produce executions with larger inconsistencies or anomalous numerical
values~\cite{guo2020audee}.  COMET expands the mutation space and guides
iterative mutation using coverage of layer inputs, parameter values, and layer
sequences~\cite{li2023comet}.  More recently, DevMuT derives mutation
operators and legality constraints from developer expertise, uses Double-Q
learning to select seed models and operators, and examines multiple stages
of the model lifecycle~\cite{mu2024devmut}.  These methods share a common
workflow: they select a seed model, a mutation operator, and a site; generate
and execute the mutant; and use an oracle and execution feedback to identify
failures or guide later mutations~\cite{mu2026howfar,zou2023ramos,zou2026dlmmm}.

\parlabel{Comparison with prior work.}
\tool uses a profiled whole-model \tfg for mutation construction, legality
checking, and candidate selection.  First, compared with API-level
methods that vary calls and execution contexts, it mutates complete executable
models and observes tensor states across operator connections.
Second, among model-level methods, it follows mutation rather than
generation from templates or operator specifications.  Prior mutation methods
primarily organize edits around layers, parameters, weights, or structural graph
units; \tool instead reprofiles each accepted mutant and uses its \tfg to form
and filter site-specific candidates using runtime tensor states and operator
input requirements.  It selects retained candidates using coverage and impact
features.  Section~\ref{sec:method} presents the design.

\section{Method}
\label{sec:method}

\subsection{Overview}

\begin{figure}[!t]
  \centering
  \includegraphics[width=\linewidth]{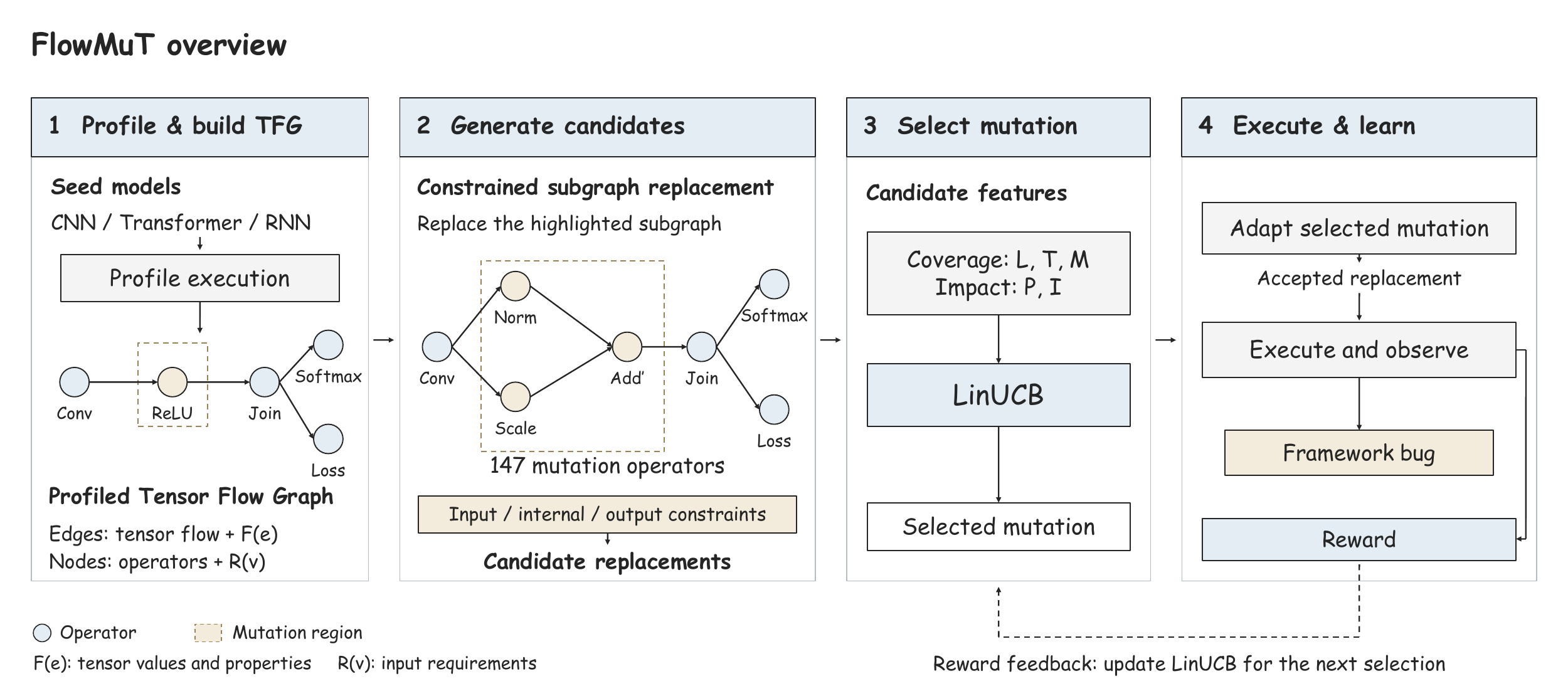}
  \caption{The \tool workflow.}
  \Description{Four stages are arranged from left to right and continue along
  a lower feedback loop.  A seed model is profiled into a TFG, a candidate is
  selected and applied to obtain a mutant, and the execution outcome guides the
  next iteration.}
  \label{fig:flowmut-overview}
\begin{minipage}{\linewidth}
\begin{algorithm}[H]
    \captionsetup{hypcap=false}
    \caption{The \tool Algorithm}
    \phantomsection
    \label{alg:flowmut}
    \footnotesize
    \algrenewcommand{\algorithmiccomment}[1]{%
      \hfill\textcolor{blue}{\(\triangleright\)~#1}}
    \begin{algorithmic}[1]
    \Function{FlowMuT}{$f,X_p,\mathcal{M},N$}
      \State $E \gets \Call{ProfileExecution}{f,X_p}$
             \Comment{1. Profile Execution}
      \State $G \gets \Call{BuildTFG}{E}$;
             $H,B \gets \emptyset$
             \Comment{2. Build the Profiled TFG}
      \For{$t \gets 1$ \textbf{to} $N$}
        \State $C \gets \Call{GenerateCandidates}{G,\mathcal{M},X_p}$
        \State \textbf{if} $C=\emptyset$ \textbf{ then break}
        \State $c \gets \Call{SelectWithLinUCB}{C,G,H}$
               \Comment{3. Generate and Select Mutations}
        \State $f' \gets \Call{Apply}{f,c}$;
               $z \gets \Call{ExecuteAndObserve}{f',X_p}$
               \Comment{4. Detection and Iterative Mutation}
        \State $H \gets \Call{UpdateLinUCB}{H,c,z}$
        \State \textbf{if} $z$ reports a bug \textbf{ then}
               $B \gets B\cup\{z\}$
        \State \textbf{else if} $\Call{SuccessfulExecution}{z}$ \textbf{ then}
               $f \gets f'$;
               $G \gets \mathrm{BuildTFG}(\mathrm{ProfileExecution}(f,X_p))$
      \EndFor
      \State \Return $B$
    \EndFunction
    \end{algorithmic}
\end{algorithm}
\end{minipage}
\end{figure}

Figure~\ref{fig:flowmut-overview} and Algorithm~\ref{alg:flowmut} summarize the
four-stage workflow.  Given a seed model $f$, profiling inputs $X_p$, and a
preconstructed operator set $\mathcal{M}$, \tool profiles the execution and
builds its \tfg.  It instantiates operators as constrained subgraph
replacements, filters illegal candidates, and uses LinUCB to select an operator
and site.  It then executes the mutant, checks for framework bugs, updates the
selection state, and refreshes the \tfg after an accepted mutation.  The
following subsections define the \tfg and mutation abstraction
(Section~\ref{sec:tfg}), construct the operator pool
(Section~\ref{sec:mutation-operators}), and present candidate construction and
selection (Section~\ref{sec:mutation-selection}).

\subsection{Tensor Flow Graph and Mutation Abstraction}
\label{sec:tfg}

\subsubsection{Tensor Flow Graph}

\begin{figure}[!t]
  \centering
  \includegraphics[width=\linewidth]{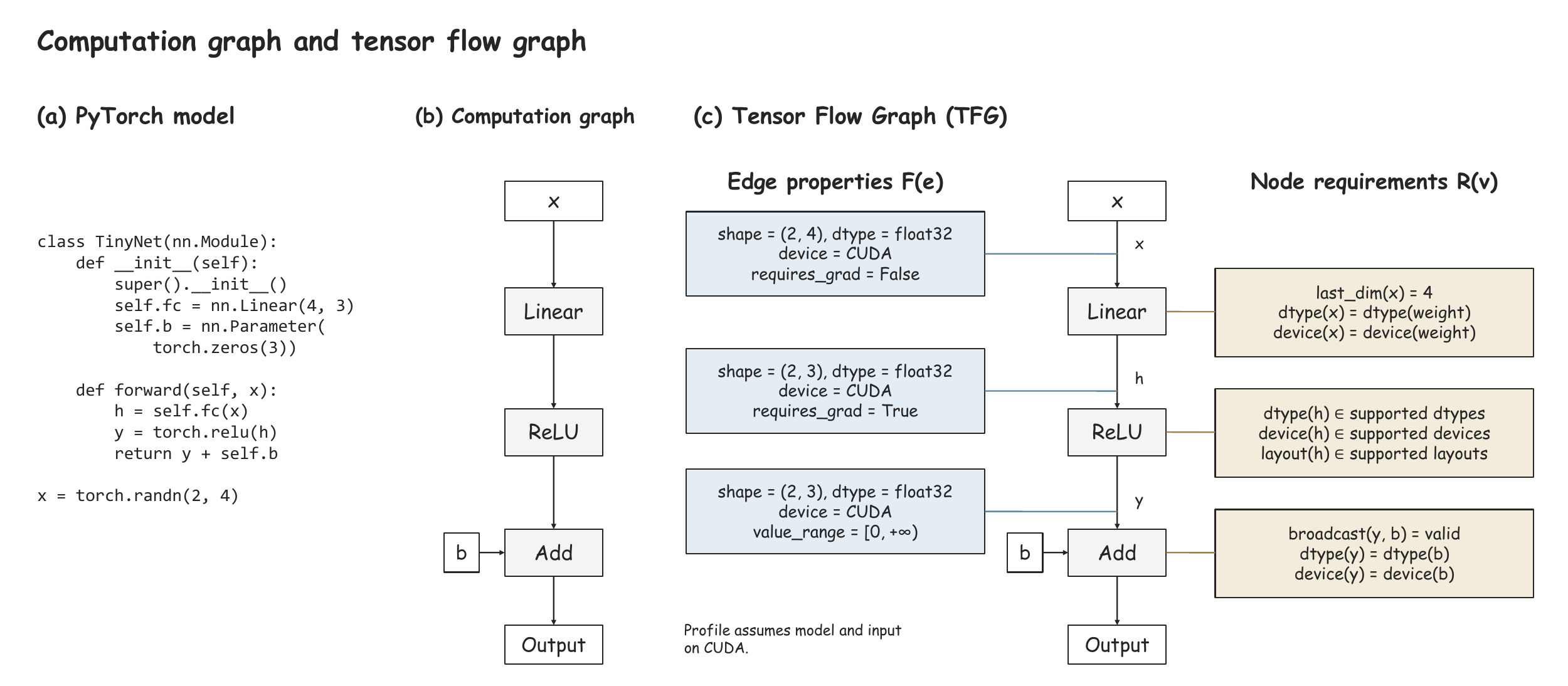}
  \caption{Construction of a \tfg from a PyTorch model, combining operator
  topology $G$, runtime tensor values and properties $F$, and operator input
  requirements $R$.}
  \Description{Three panels show the PyTorch source model, its computation
  graph, and the resulting TFG.  The TFG preserves operator connections,
  annotates tensor edges with runtime values and properties, and attaches input
  requirements to consumer operators.}
  \label{fig:tfg}
\end{figure}

A conventional graph captures operator topology but not the runtime tensor
values and properties or operator input requirements needed for data flow
mutation.  \tool brings these three kinds of information together in a Tensor
Flow Graph:
\[
  \mathrm{TFG}=(G,F,R),
  \qquad G=(V,E).
\]
Figure~\ref{fig:tfg} illustrates this description using a PyTorch model.

Here, $G$ records the operator topology of the profiled execution.  Each node
$v\in V$ represents an operator, and each directed edge $e\in E$ represents a
tensor flowing from its producer to one or more consumers.  We use $U(e)$ to
denote the direct consumers of $e$.  This operator topology identifies where a
tensor is created, how it moves through the model, and which downstream
operators may be affected when it is changed.

The edge annotation $F(e)$ describes the tensor carried by edge $e$.  \tool
collects these annotations by profiling the model with inputs $X_p$.  They
include tensor values and value summaries, shape, rank, dtype, device, memory
layout, aliasing information, numerical properties, and gradient state.  When
profiling inputs produce different properties on an edge, $F(e)$ keeps the
states needed to check whether a mutation operator can be applied across
those inputs.  The node annotation $R(v)$ records the operator input
requirements associated with $v$.  These requirements include properties of
individual inputs, such as rank, dtype, device, layout, gradient state, and
valid value range, as well as relationships among inputs, such as dimension
compatibility, broadcasting, and device agreement.  We construct $R(v)$ by
extracting operator input requirements from the official documentation and
source repositories of the supported frameworks.  For each operator, we
normalize the extracted requirements into checks on individual input tensors
and relations among multiple inputs, and attach the checks to the
corresponding node annotation.  The profiled tensor states in $F(e)$ provide
the concrete values against which these checks are evaluated; they do not
define the requirements themselves.  Thus, $F(e)$ describes the tensor in
the data flow, whereas $R(v)$ describes the tensors that a consumer can
accept.

Thus, $G$ identifies mutation sites, $F$ determines which mutation operators
can be instantiated, and $R$ checks their compatibility with the surrounding
model.

\subsubsection{Mutation as Constrained Subgraph Replacement}

To unify mutations over tensors, operators, and bounded computations, \tool
represents each mutation as the constrained replacement of one bounded
subgraph with another.  Each mutation operator has three parts: a subgraph
pattern, a replacement, and a set of constraints.  The pattern states where
the operator can be applied, the replacement states how the matched subgraph
is changed, and the constraints state when the change is expected to remain
legal.

\begingroup\looseness=-1
Each mutation operator has input, internal, and output constraints.  Input
constraints check edges produced outside the matched subgraph and consumed
inside it.  Internal constraints check the topology and data flow within the
matched and replacement subgraphs.  Output constraints check edges produced
inside the replacement and consumed by nodes outside it.  Together, these
constraints stop \tool from applying mutations that are incompatible with the
surrounding model.\par\endgroup

\subsection{Mutation Operator Pool}
\label{sec:mutation-operators}

\begin{figure}[!t]
  \centering
  \includegraphics[width=\linewidth]{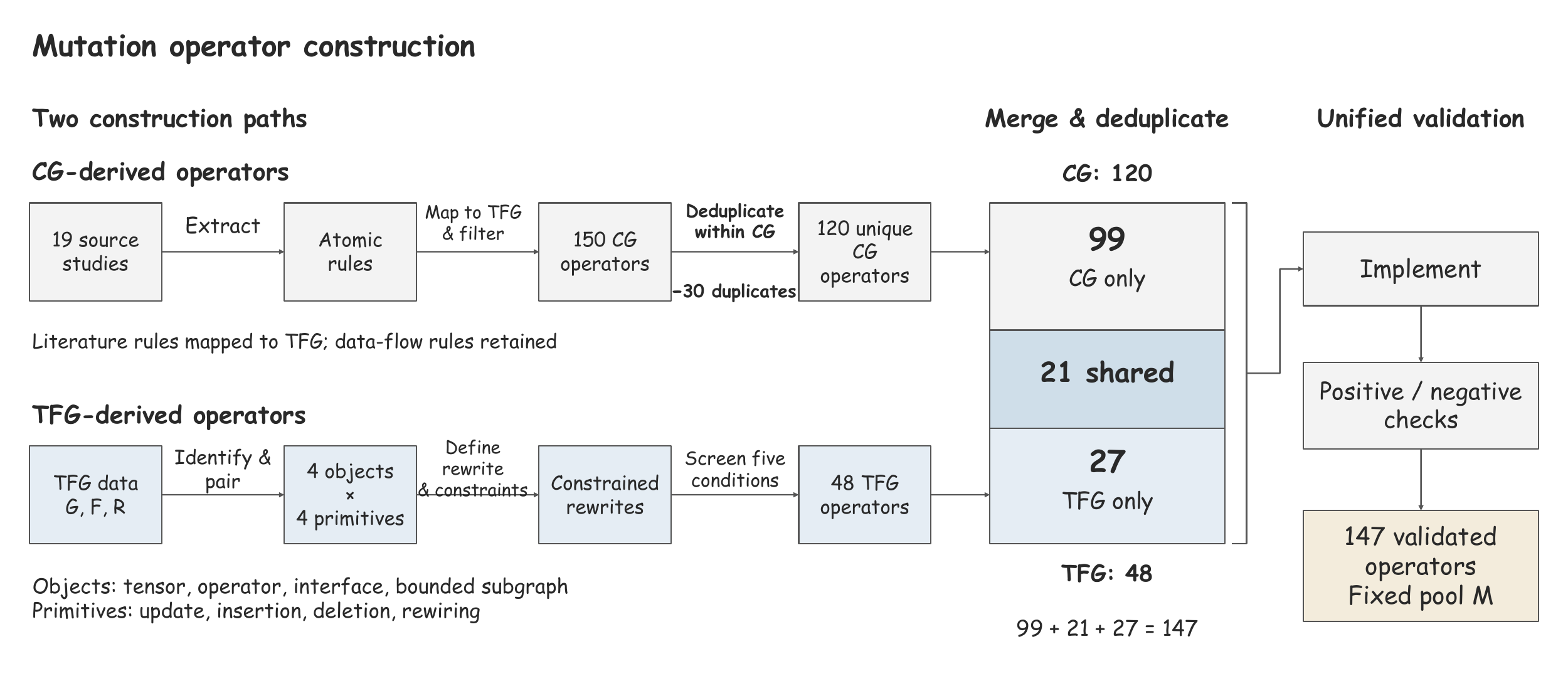}
  \caption{Construction of the \tool mutation operator pool from CG-derived and
  TFG-derived operators.}
  \Description{The two sources produce 120 unique CG-derived operators and 48
  TFG-derived operators, of which 21 are shared.  Their union is implemented
  and validated into a pool of 147 operators.}
  \label{fig:operator-pool}
\end{figure}

\tool combines CG-derived operators adapted from prior work with TFG-derived
operators constructed from mutable TFG objects and mutation primitives.  It
then deduplicates and validates their union, as summarized in
Figure~\ref{fig:operator-pool}.

\subsubsection{CG-Derived Mutation Operators}
\label{sec:operator-derivation}

\begingroup\looseness=-1
The CG-derived operators adapt concrete mutation rules defined over CGs in
prior DL framework testing studies to the TFG-based description.  This process
proceeds in four steps.  \textbf{First,} we select source studies.  We begin
with the studies discussed in Section~\ref{sec:related}, follow their
references and later citing studies, and stop when a complete round yields no
new study that meets our inclusion rule.  We include a study only when it
defines at least one concrete and reproducible change to a tensor, a framework
operator, or a bounded computation.  This procedure identifies 19 source
studies:
\cite{wang2020lemon,guo2020audee,li2023comet,mu2024devmut,
mu2026howfar,zou2024dljsfuzzer,liu2024gandalf,wang2025d3,
liu2023neuri,wang2022eagle,zhang2025orthrus,wei2022freefuzz,
deng2023titanfuzz,ji2025moco,li2026dllens,
christou2023ivysyn,liu2022tzer,su2023torchprobe,ma2023hirgen}.
\textbf{Second,} we extract the mutation rules from each selected study and
record their source locations, targets, actions, parameters, preconditions,
and intended effects.  When one rule contains independent edits, we split it
into atomic rules, while different values of the same parameter remain one
parameterized rule.  \textbf{Third,} we map each extracted rule to the TFG.  We
retain a rule only when its target maps to a TFG edge, node, or bounded
subgraph, its edit changes the topology or a recorded tensor or operator
property, and the information needed to check it is available from the TFG or
a documented framework requirement.  Each retained rule is expressed as a
mutation operator with a pattern, a replacement, and input, internal, and
output constraints.  \textbf{Fourth,} we remove rules that describe only a
testing strategy or change framework internals without changing model-level
data flow.  We extract 155 rules from the 19 studies.  Mapping and filtering
remove five rules, and deduplication within the CG-derived set removes 30 more,
leaving 120 unique CG-derived mutation operators.\par\endgroup

\subsubsection{TFG-Derived Mutation Operators}
\label{sec:tfg-guided-operator-synthesis}

The TFG-derived operators complement prior topology-oriented rules with
mutations based on additional data flow information.  This process proceeds in
four steps.  \textbf{First,} using $G$, $F$, and $R$, we identify four types of
mutable objects: tensors and their states, operators, producer--consumer
interfaces and input relations, and bounded data flow structures.
\textbf{Second,} we define four basic ways to change a mutable object, which we
call \emph{mutation primitives}.  An \emph{update} changes the
value or label of an object without changing its connections.  An
\emph{insertion} places a new operator and its connected edges on a tensor edge
or at an interface.  A \emph{deletion} removes a node or bounded subgraph and
reconnects its boundary.  A \emph{rewiring} changes a producer--consumer
connection within a bounded region.

\parlabel{Third,} we combine each mutable object type with each mutation
primitive and retain the meaningful combinations, each of which may define one
or more transformations.  Each resulting operator specifies the target object,
the primitive and its parameters, the properties that must remain unchanged,
and the TFG feature that should change.  These elements determine the subgraph
pattern and replacement, to which we add the input, internal, and output
constraints defined in Section~\ref{sec:tfg}.
\textbf{Fourth,} we check each derived operator against five conditions.  Its
target and the information needed to check it must be available in $G$, $F$,
or $R$, or in a documented framework requirement.  The primitive must change
at least one recorded state or relation, the changed region must have a clear
boundary, supported framework operations must be able to implement the
replacement, and the replacement must satisfy the operator input requirements
at its boundaries.  This process produces 48 reusable TFG-derived mutation
operators.  Concrete values and model locations are assigned later when an
operator is matched to a profiled TFG.

\begin{table*}[!t]
  \caption{Construction of the 147-operator pool by mutable TFG object,
  mutation primitive, and source.}
  \label{tab:operator-construction}
  \centering
  \small
  \setlength{\tabcolsep}{4.5pt}
  \renewcommand{\arraystretch}{0.92}
  \begin{tabular}{ll@{\hspace{0.9em}}rrr@{\hspace{0.9em}}rrr@{\hspace{0.9em}}>{\columncolor{tableunion!55}[0pt][\tabcolsep]}r}
    \toprule
    \multicolumn{2}{>{\columncolor{tablecanonical}[\tabcolsep][0pt]}c}{\textbf{Mutation category}} &
    \multicolumn{3}{>{\columncolor{tablecg}[0pt][0pt]}c}{\textbf{CG-derived construction}} &
    \multicolumn{3}{>{\columncolor{tabletfg}[0pt][0pt]}c}{\textbf{TFG-derived construction}} &
    \multicolumn{1}{>{\columncolor{tableunion}[0pt][\tabcolsep]}c}{\textbf{Final pool}} \\
    \multicolumn{1}{>{\columncolor{tablecanonical}[\tabcolsep][\tabcolsep]}l}{Object} &
    \multicolumn{1}{>{\columncolor{tablecanonical}[\tabcolsep][0pt]}l}{Primitive} &
    \multicolumn{1}{>{\columncolor{tablecg}[0pt][\tabcolsep]}c}{Extracted} &
    \multicolumn{1}{>{\columncolor{tablecg}[\tabcolsep][\tabcolsep]}c}{Removed} &
    \multicolumn{1}{>{\columncolor{tablecg}[\tabcolsep][0pt]}c}{Retained} &
    \multicolumn{1}{>{\columncolor{tabletfg}[0pt][\tabcolsep]}c}{Derived} &
    \multicolumn{1}{>{\columncolor{tabletfg}[\tabcolsep][\tabcolsep]}c}{Overlap} &
    \multicolumn{1}{>{\columncolor{tabletfg}[\tabcolsep][0pt]}c}{TFG-only} &
    \multicolumn{1}{>{\columncolor{tableunion}[0pt][\tabcolsep]}c}{Total} \\
    \midrule
    Tensor & Update    & 19 & 14 & 5 & 8 & 5 & 3 & 8 \\
    Tensor & Insertion &  1 &  0 & 1 & 4 & 1 & 3 & 4 \\
    \rowcolor{tablesubtotal}
    \multicolumn{2}{r}{\textbf{Tensor subtotal}} &
      \textbf{20} & \textbf{14} & \textbf{6} &
      \textbf{12} & \textbf{6} & \textbf{6} & \textbf{12} \\
    \addlinespace[2pt]
    Operator & Update    & 46 & 9 & 37 & 6 & 5 & 1 & 38 \\
    Operator & Insertion &  6 & 3 &  3 & 3 & 2 & 1 &  4 \\
    Operator & Deletion  &  2 & 1 &  1 & 1 & 1 & 0 &  1 \\
    \rowcolor{tablesubtotal}
    \multicolumn{2}{r}{\textbf{Operator subtotal}} &
      \textbf{54} & \textbf{13} & \textbf{41} &
      \textbf{10} & \textbf{8} & \textbf{2} & \textbf{43} \\
    \addlinespace[2pt]
    Interface & Update    & 49 & 6 & 43 & 2 & 1 & 1 & 44 \\
    Interface & Insertion &  0 & 0 &  0 & 6 & 0 & 6 &  6 \\
    Interface & Deletion  &  1 & 0 &  1 & 2 & 0 & 2 &  3 \\
    Interface & Rewiring  &  0 & 0 &  0 & 3 & 0 & 3 &  3 \\
    \rowcolor{tablesubtotal}
    \multicolumn{2}{r}{\textbf{Interface subtotal}} &
      \textbf{50} & \textbf{6} & \textbf{44} &
      \textbf{13} & \textbf{1} & \textbf{12} & \textbf{56} \\
    \addlinespace[2pt]
    Subgraph & Update    & 16 & 1 & 15 & 4 & 2 & 2 & 17 \\
    Subgraph & Insertion &  9 & 0 &  9 & 3 & 2 & 1 & 10 \\
    Subgraph & Deletion  &  1 & 0 &  1 & 2 & 1 & 1 &  2 \\
    Subgraph & Rewiring  &  5 & 1 &  4 & 4 & 1 & 3 &  7 \\
    \rowcolor{tablesubtotal}
    \multicolumn{2}{r}{\textbf{Subgraph subtotal}} &
      \textbf{31} & \textbf{2} & \textbf{29} &
      \textbf{13} & \textbf{6} & \textbf{7} & \textbf{36} \\
    \midrule
    \rowcolor{tablegrand}
    \multicolumn{2}{r}{\textbf{Grand total}} &
      \textbf{155} & \textbf{35} & \textbf{120} &
      \textbf{48} & \textbf{21} & \textbf{27} & \textbf{147} \\
    \bottomrule
  \end{tabular}
\end{table*}

This approach systematically generates mutation operators related to data
flow.  For example, combining an alias-annotated tensor edge with the insertion
primitive produces an operator that inserts a copy and breaks sharing while
preserving the tensor interface.  Combining a shared edge with the rewiring
primitive produces a branch-local operator that changes only one consumer
path.  Combining a producer--consumer interface with the insertion primitive
produces an operator that makes an implicit dtype or layout conversion
explicit.

\subsubsection{Merging, Deduplication, and Validation}

In the unified TFG representation, two operators are duplicates when they have
the same target, edit, parameters, and applicability conditions.  Deduplicating
the 120 CG-derived and 48 TFG-derived operators identifies 21 overlaps and
yields the 147-operator pool summarized in
Table~\ref{tab:operator-construction}.
For each operator in this union, we define an implementation using an anchor
edge, a TFG pattern, a replacement, and input, internal, and output
constraints.  We adapt the implementation to each supported framework while
preserving the operator's target, effect, and constraints.  We retain it only
when positive and negative checks confirm that it performs the intended change
and rejects violations of its declared constraints.  This process produces
the fixed pool $\mathcal{M}$ of 147 validated \tool mutation operators, which
we fix before running the evaluation.

\subsection{Test Case Generation}
\label{sec:mutation-selection}

In each iteration, \tool generates a framework test case by mutating the
current seed model and executing the resulting mutant.  It first profiles the
seed model $f$ to obtain a TFG.  Using this TFG, it pairs mutation operators
with sites to construct candidates and filters those that violate data flow
constraints.  From the retained candidates, it selects one using coverage and
impact features and applies it to $f$ to obtain a mutant $f'$.  It executes
$f'$ and, if the execution succeeds, uses it as the seed model for the next
iteration.  The following subsections detail how these candidates are
constructed and selected.

\subsubsection{Efficient Candidate Construction}
\label{sec:candidate-construction}

A candidate $c=(s,m)$ pairs a mutation operator $m$ with a TFG site $s$
anchored at a tensor edge.  Rather than enumerating every subgraph, \tool
visits each tensor edge $e\in E$, treats it as the anchor of a possible site,
and pairs the site with every mutation operator.  Because the mutation operator
pool is fixed before testing, the work for one edge is constant, giving
candidate construction a time complexity of $O(|E|)$.

After candidate construction, \tool filters the candidates using the input,
internal, and output constraints.  The internal constraints determine whether
the data flow around the anchor edge matches the pattern $p$ of $m$, while the
input and output constraints check the two boundaries of the mutation site.  A
candidate is retained only when it satisfies all three types of constraints.
The retained candidates form $\mathcal{C}_t$ for the selection procedure
described next.

\subsubsection{Candidate Selection}
\label{sec:candidate-selection}

\parlabel{Coverage features.}
We define three coverage features to represent the site context, tensor state,
and mutation operator of a mutation candidate.  The site context feature uses
the anchor, the operators, and the tensor edges in $s$.
The tensor state feature uses rank, dtype, device, layout, gradient state, and
numerical properties from $F$.  The mutation operator feature records the
mutation operator and its transformation.  For each feature set $B$, \tool
computes its coverage value as
\[
  d(B)
  =
  \frac{1}{|B|}
  \sum_{b\in B}
  \frac{1}{\sqrt{1+n(b)}},
\]
where $n(b)$ is the number of previously executed mutation candidates
containing element $b$.  Thus, an unseen element contributes $1$, and its
contribution decreases as it is exercised repeatedly.  Applying $d$ to the
three feature sets gives the site context, tensor state, and mutation operator
coverage features $L(c)$, $T(c)$, and $M(c)$, respectively.

\parlabel{Impact features.}
We define two impact features to represent how broadly and strongly a mutation
candidate may affect the executed model.  The downstream impact feature
measures the fraction of executed computation reachable from the output
boundary of the subgraph replaced by candidate $c$.  For a profiling input
$x$, let $D(c,x)$ contain the executed operators downstream of that subgraph,
and let $V(x)$ contain all operators executed by $x$.  \tool computes this
feature by averaging the reachable fraction over the profiling inputs:
\[
  P(c)
  =
  \frac{1}{|X_p|}
  \sum_{x\in X_p}
  \frac{|D(c,x)|}{|V(x)|}.
\]
The reachability sets are computed directly from $G$ and cached for candidates
that share the same output boundary.  The gradient influence feature measures
how strongly the model output depends on candidate $c$ through its anchor
tensor.  Let $a(c,x)$ be the runtime tensor on the anchor edge of $s$, and let
$\ell$ reduce the model output to a scalar.  \tool computes the average
normalized absolute gradient:
\[
  g(c)
  =
  \frac{1}{|X_p|}
  \sum_{x\in X_p}
  \frac{1}{|a(c,x)|}
  \left|
  \frac{\partial \ell(f(x))}{\partial a(c,x)}
  \right|.
\]
Dividing by the number of tensor elements prevents tensor size from dominating
the feature value.  For executions without gradients, \tool estimates influence by
perturbing the anchor tensor and measuring the output response.  It normalizes
$g(c)$ to $[0,1]$ within the current candidate set and denotes the result by
$I(c)$.

Together, the three coverage features $L(c)$, $T(c)$, and $M(c)$ and the two
impact features $P(c)$ and $I(c)$ form the context vector for each candidate.

\parlabel{Selection with LinUCB.}
A candidate's value is observable only after
execution~\cite{mu2022dmos,mu2021hars}, while each accepted
mutation changes the TFG and hence the candidate set.  We therefore model
candidate selection as a contextual bandit problem.  The five features above
describe each candidate, and the reward measures the novelty of the resulting
framework behavior.  \tool uses LinUCB~\cite{li2010contextual}
because its linear estimate transfers feedback among candidates with similar
contexts as their identities change, while its upper-confidence term encourages
the selection of uncertain candidates.  At iteration $t$, LinUCB scores every
$c\in\mathcal{C}_t$ by its estimated reward and contextual uncertainty, then
selects the highest-scoring candidate.

After execution, \tool records the mutant's structured runtime trace
$\tau_t=(e_1,\ldots,e_K)$.  A framework adapter maps dispatcher choices,
autograd actions, graph-capture and optimization decisions, tensor adaptation,
memory or alias handling, and synchronization to control-flow labels.  It
excludes tensor values and properties, producer or consumer identities, and
kernel names, keeping the trace distinct from the tensor-level signatures used
by DFSD.  Let $E_t$ and $Q_t$ be the event labels and adjacent transitions in
$\tau_t$, and let $E_{<t}=\bigcup_{i<t}E_i$ and $Q_{<t}=\bigcup_{i<t}Q_i$
denote those observed in earlier executions.  The reward is
\[
  r_t =
  \frac{1}{2}\frac{|E_t\setminus E_{<t}|}{\max(1,|E_t|)}
  +
  \frac{1}{2}\frac{|Q_t\setminus Q_{<t}|}{\max(1,|Q_t|)}.
\]
Thus, $r_t$ equally weights the fractions of newly observed events and
transitions.  Traces with no new behavior or no complete execution receive
zero, and failure labels are excluded.  \tool then updates the event,
transition, and coverage histories and LinUCB state with the candidate context
and $r_t$.  In contrast,
DFSD is an offline metric over tensor-level signatures in a common test space;
$r_t$ measures newly exposed framework behavior online per execution.

Each accepted mutant becomes the next seed model.  Its execution is reprofiled
to refresh the TFG and candidate set.

\section{Experiment Design}
\label{sec:experiment-design}

\subsection{Research Questions}

This study investigates the following research questions:

\begin{itemize}
  \item \textbf{RQ1: Bug Detection.} How effective is \tool in
  detecting real bugs in DL frameworks, and what types of bugs does it uncover?
  \item \textbf{RQ2: Diversity and Coverage.} How effectively does
  \tool improve test case diversity and framework coverage?
  \item \textbf{RQ3: Mutation Legality.} How effectively do
  TFG-based constraints filter illegal mutations?
  \item \textbf{RQ4: Mutation Selection.} How effectively does
  LinUCB guide the joint selection of mutation operators and sites?
\end{itemize}

\subsection{Benchmarks}

\begingroup\looseness=-1
\textbf{DL frameworks and models.} The evaluation targets PyTorch in eager and
compiled modes and MindSpore in PyNative and Graph modes.
Section~\ref{sec:implementations} reports the framework versions, devices, and
execution environments.  The benchmark contains 14 seed models spanning ten
application tasks, from image classification and dense prediction to language
and image generation.  The selected architectures are recent and
representative, and compact or small variants serve as the seed models when
several scales are available~\cite{woo2023convnextv2,liu2022swinv2,tan2021efficientnetv2,
wang2024yolov10,lv2024rtdetrv2,xie2021segformer,cheng2022mask2former,
warner2025modernbert,batzner2024efficientad,
jiang2023rtmpose,bautista2022parseq,qwen2024qwen25,
peebles2023dit,liang2021swinir}.
Table~\ref{tab:setup} summarizes the benchmark, and the artifact records the
model sources, checkpoints, profiling inputs, and checksums.\par\endgroup

\begin{table}[!t]
\caption{Seed models by application task, input size, and parameter count.}
\label{tab:setup}
\centering
\fontsize{8.5pt}{10pt}\selectfont
\setlength{\tabcolsep}{2.5pt}
\renewcommand{\arraystretch}{0.96}
\begin{tabular}{c c c r}
\toprule
\rowcolor{tableheader}
\textbf{Scene Task} & \textbf{Model} & \textbf{Input} & \textbf{Parameters} \\
\midrule
\multirow[c]{3}{*}{Image Classification}
 & ConvNeXt V2-Tiny & 224$\times$224 & 28,635,496 \\
 & Swin V2-Tiny & 256$\times$256 & 28,347,154 \\
 & EfficientNetV2-S & 384$\times$384 & 21,458,488 \\
\midrule
\multirow[c]{2}{*}{Object Detection}
 & YOLOv10-S & 640$\times$640 & 8,128,272 \\
 & RT-DETRv2-S & 640$\times$640 & 20,226,500 \\
\midrule
\multirow[c]{2}{*}{Semantic Segmentation}
 & SegFormer-B1 & 512$\times$512 & 13,715,798 \\
 & Mask2Former-Swin-T & 512$\times$512 & 47,441,169 \\
\midrule
Text Classification & ModernBERT-base & 1$\times$512 & 149,606,402 \\
\midrule
Industrial Anomaly Detection & EfficientAD-S & 256$\times$256 & 8,057,856 \\
\midrule
Keypoint Detection & RTMPose-m & 256$\times$192 & 13,587,611 \\
\midrule
Scene Text Recognition & PARSeq & 32$\times$128 & 23,832,671 \\
\midrule
Text Generation & Qwen2.5-0.5B & 1$\times$512 & 494,032,768 \\
\midrule
Image Generation & DiT-S/2 & 256$\times$256 & 32,963,360 \\
\midrule
Image Restoration & SwinIR-M ($\times4$) & 256$\times$256 LR & 11,900,199 \\
\bottomrule
\end{tabular}
\end{table}

\subsection{Baseline Methods}

LEMON, COMET, and DevMuT serve as baselines because they generate model-level
test cases through mutation~\cite{wang2020lemon,li2023comet,mu2024devmut}.
Following their papers and released implementations, we adapt all three to the
same PyTorch and MindSpore versions used by \tool.

\subsection{Measurements}

\parlabel{RQ1: Bug detection.} We count distinct reproducible bugs
reported to framework developers and record whether they are confirmed or
fixed.

\parlabel{Test-case legality and oracle.} A test case is legal when its
mutation satisfies the mutation specification and the documented preconditions
of the affected operators, independently of its execution outcome.  An
execution failure caused by a violation of these conditions indicates an
illegal test case.  For a legal test case, the oracle identifies a potential
framework bug when execution unexpectedly crashes or raises an uncaught
exception, violates documented semantics, or produces a difference beyond the
expected numerical tolerance across semantically equivalent execution modes.
We count a failure as a bug only when it can be reproduced.

\parlabel{RQ2: Diversity and coverage.} We use three established
metrics, Layer Input Coverage (LIC), Layer Parameter Coverage (LPC), and Layer
Sequence Coverage (LSC), together with our Data Flow Signature Diversity
(DFSD)~\cite{wang2020lemon,gu2022muffin,li2023comet,liu2023nnsmith,
mu2026howfar,xie2026sort}.  LIC measures covered input datatypes,
dimensionalities, and shapes; LPC measures covered parameter values; and LSC
measures covered valid producer--consumer operator pairs.  DFSD measures
covered data flow signatures, where each signature records the producer and
consumer operator types, operator or kernel path, binned value profile, and
tensor properties, including shape, rank, dtype, layout, device, aliasing, and
gradient state.  The four metrics are computed as
\mbox{$\mathrm{LIC} =
\frac{N_{\text{covered input values}}}
     {N_{\text{input values in the test space}}}$},
\mbox{$\mathrm{LPC} =
\frac{N_{\text{covered parameter values}}}
     {N_{\text{parameter values in the test space}}}$},
\mbox{$\mathrm{LSC} =
\frac{N_{\text{covered valid pairs}}}{N_{\text{valid pairs}}}$}, and
\mbox{$\mathrm{DFSD} =
\frac{N_{\text{observed distinct signatures}}}
     {N_{\text{signatures in the test space}}}$}.
The LIC value is macro-averaged over operator types within each run.  We also
report the distinct signature count.  Final LIC, LPC, LSC, DFSD, and distinct
signature counts at round 250 are means across repeated runs.

\parlabel{Coverage.} We
measure Python statement coverage with \texttt{coverage.py}, C/C++ source-line
coverage with \texttt{LCOV} and \texttt{GCOV}, and API coverage as
the proportion of public interfaces included in the evaluation that are
exercised by legal test cases.  We report these three framework-coverage
measures as means across repeated runs at round 250.  For the operator
ablation studies, we also record the fixed operator-pool size.  We measure
practical cost as the average generation and detection time per test case.

\parlabel{RQ3: Mutation legality.} The primary measure is the
Legal-Mutant Rate (LMR), which is the fraction of retained candidates that
satisfy the legality definition above.  Thus,
\mbox{$\mathrm{LMR}=
\frac{N_{\text{legal retained mutants}}}{N_{\text{retained mutants}}}$.}
For each configuration, mutation attempts are the candidates retained by its
static filter.  We apply the complete constraint system and four variants that
omit the input, internal, output, or all three constraint types to the same
1,000 pre-filter candidates.  Manual inspection labels 864 as legal and 136 as
illegal according to the mutation specification and documented operator
preconditions.  We report candidate retention, LMR, and filter selectivity,
and categorize each illegal candidate by its first violated
constraint type.

\parlabel{RQ4: Mutation selection.} The analysis compares LinUCB with
one-shot, random-search, fixed-priority, and UCB1 selection.  Every policy
starts from the same seed models and initial candidate sets and uses the same
candidate-generation and legality-filtering procedures.  Each policy is
evaluated for 250 mutation rounds.
LIC, LPC, LSC, DFSD, and distinct signature counts are recorded at
fixed checkpoints to compare both final diversity and growth over time.  To
analyze the LinUCB context, each of its five features is removed in turn: site
coverage, tensor state coverage, mutation operator coverage, downstream impact,
and gradient influence.  The complete context is additionally compared with
coverage-only and impact-only variants.

\subsection{Implementations}
\label{sec:implementations}

\tool is implemented for PyTorch and MindSpore, and the released
implementations of LEMON, COMET, and DevMuT are adapted to the same framework
versions.  All methods use the same seed models, saved inputs, and execution
contexts.  We evaluate PyTorch 2.12.1 in eager and compiled modes with Python
3.13.12 and CUDA 13.0, and MindSpore 2.7.1 in PyNative and Graph modes with
Python 3.9.25.  All experiments run on Ubuntu 24.04.4 LTS with NVIDIA driver
591.86, an Intel Core i7-14700K CPU, 32\,GB of RAM, and one 24\,GB NVIDIA
GeForce RTX 4090 GPU.  The artifact lists the exact framework commits.  For
bug discovery, each method ran a 12-hour campaign for each of the 14 seed
models on each framework, giving 28 campaigns per method and 112 in total.  For
the comparative evaluation, each method or ablation configuration used 10
independent runs for every model--framework pair, with 250 mutation rounds per
run.  Each run initializes the selection state and updates it across these
rounds.  Each configuration therefore contains 280 runs and 70,000 mutation
rounds.  The baselines use the fixed hyperparameters recommended in their
papers and repositories.

\section{Results and Analysis}
\label{sec:results}

\subsection{RQ1: Bug Detection}

We compare the methods by the number of distinct bugs they detect.  For cases
found only by \tool, we report the confirmation and fix status of the submitted
bug reports, classify them by observable symptom, and analyze representative
cases in each category.

\begin{figure}[!htbp]
  \centering
  \includegraphics[width=\linewidth]{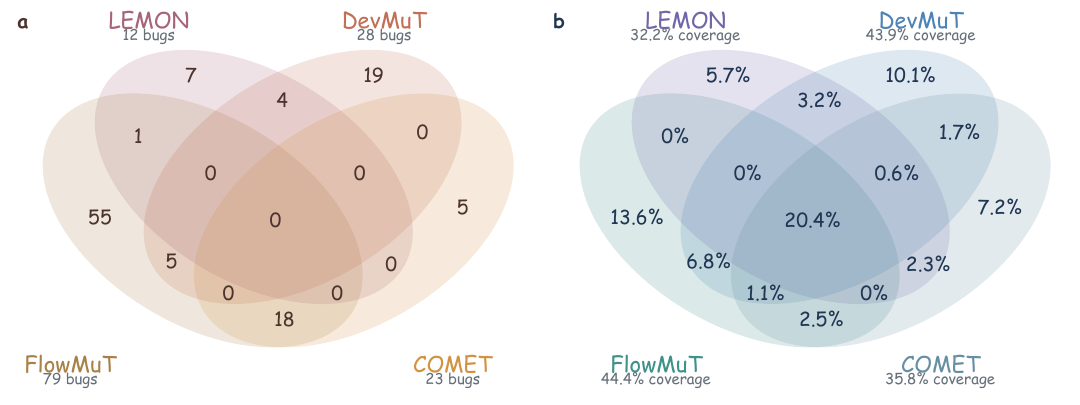}
  \caption{(a) Bug overlap from discovery campaigns; (b) C++ line overlap
  from one run per method.}
  \Description{Two four-set overlap diagrams compare LEMON, DevMuT, FlowMuT,
  and COMET.  Panel (a) shows detected-bug overlap, including 55 cases found
  only by FlowMuT and submitted as bug reports.  Panel (b) shows C++ source-line
  coverage from one run per method: 32.2 percent for LEMON, 43.9 percent for
  DevMuT, 44.4 percent for FlowMuT, and 35.8 percent for COMET.}
  \label{fig:coverage-bug-overlap}
\end{figure}

\parlabel{Detected bugs.}
We call a bug report \emph{new} when it documents a reproducible case detected
only by \tool and no similar issue previously existed in the corresponding
framework's issue tracker.
Across the bug-discovery campaigns, LEMON, COMET, DevMuT, and \tool
detected 12, 23, 28, and 79 bugs, respectively.  As
Figure~\ref{fig:coverage-bug-overlap}(a) shows, 55 cases found only by \tool
led to new bug reports: 14 for PyTorch and 41 for MindSpore.
Table~\ref{tab:bug-summary} summarizes the status of these reports and groups
them by primary observable symptom.  Numerical errors include incorrect finite
results and unexpected NaN/Inf values.

\begin{table}[!htbp]
  \caption{Cumulative developer status and symptom categories of 55 new bug
  reports.}
  \label{tab:bug-summary}
  \centering
  \footnotesize
  \setlength{\tabcolsep}{2.5pt}
  \renewcommand{\arraystretch}{0.96}
  \begin{tabular}{P{0.13\linewidth}C{0.125\linewidth}C{0.135\linewidth}@{\hspace{0.6em}}P{0.23\linewidth}C{0.125\linewidth}C{0.135\linewidth}}
    \toprule
    \rowcolor{tableheader}
    \multicolumn{3}{c}{\textbf{Developer status}} &
      \multicolumn{3}{c}{\textbf{Symptom category}} \\
    \cmidrule(lr){1-3}\cmidrule(lr){4-6}
    \rowcolor{tableheader}
    \textbf{Status} & \textbf{PyTorch} & \textbf{MindSpore} &
      \textbf{Category} & \textbf{PyTorch} & \textbf{MindSpore} \\
    \midrule
    Reported & 14 (100.0\%) & 41 (100.0\%) &
      Crash & 0 (0.0\%) & 31 (75.6\%) \\
    Confirmed & 8 (57.1\%) & 33 (80.5\%) &
      Numerical Error & 12 (85.7\%) & 6 (14.6\%) \\
    Fixed & 7 (50.0\%) & 0 (0.0\%) &
      State Error & 2 (14.3\%) & 4 (9.8\%) \\
    \bottomrule
  \end{tabular}
\end{table}

\parlabel{Crash reports.}
All 31 reports concerned MindSpore.  They mainly describe cases in which
calls legal in PyNative mode or allowed by the documentation failed during
Graph-mode compilation, CPU-kernel dispatch, or argument validation.
Representative causes included missing or mismatched Tensor-method
registrations, rejection of documented scalar arguments, incomplete
sparse-CSR paths, and control-flow compilation failures.

\parlabel{Numerical error reports.}
The 18 reports describe silent changes to values or gradients.  Six concerned
MindSpore and 12 PyTorch.  The MindSpore reports primarily involved lost float64
precision or inconsistent PyNative/Graph semantics, whereas the PyTorch reports
describe compiled-execution cases involving validation bypasses, incorrect
gradients, integer overflow, and five NaN/Inf errors.  For example, a
no-argument \texttt{Tensor.argmax} returned a scalar in PyNative mode but a
vector in Graph mode, while compiled \texttt{torch.acosh} converted finite
inputs to infinity.

\begingroup\looseness=-1
\parlabel{State error reports.}
Six reports describe silent changes in dtype, shape
metadata, return structure, or RNG state.  Examples include Graph-mode
\texttt{Tensor.max/min} dropping indices and compiled
\texttt{fractional\_max\_pool2d} consuming RNG
state differently from eager execution.\par\endgroup

Analysis of a MindSpore crash detected by \tool revealed two
denial-of-service (DoS) vulnerabilities.  MindSpore's security team
acknowledged both findings.

\rqanswer{RQ1}{\tool detected 79 bugs, compared with 28 for the strongest
baseline.  It submitted 55 new bug reports for cases missed by the baselines;
developers confirmed 41 reports and fixed the bugs described in 7 reports.}

\subsection{RQ2: Diversity and Coverage}

We first compare \tool with the baselines and then examine the contributions
of its mutation operators.  LIC, LPC, and LSC characterize the diversity of
input properties, parameter values, and connected operator pairs, while DFSD
measures the diversity of data flow signatures.  Framework coverage indicates
how much implementation code and how many APIs the generated tests exercise.

\begin{figure}[!tb]
  \centering
  \includegraphics[width=0.95\linewidth]{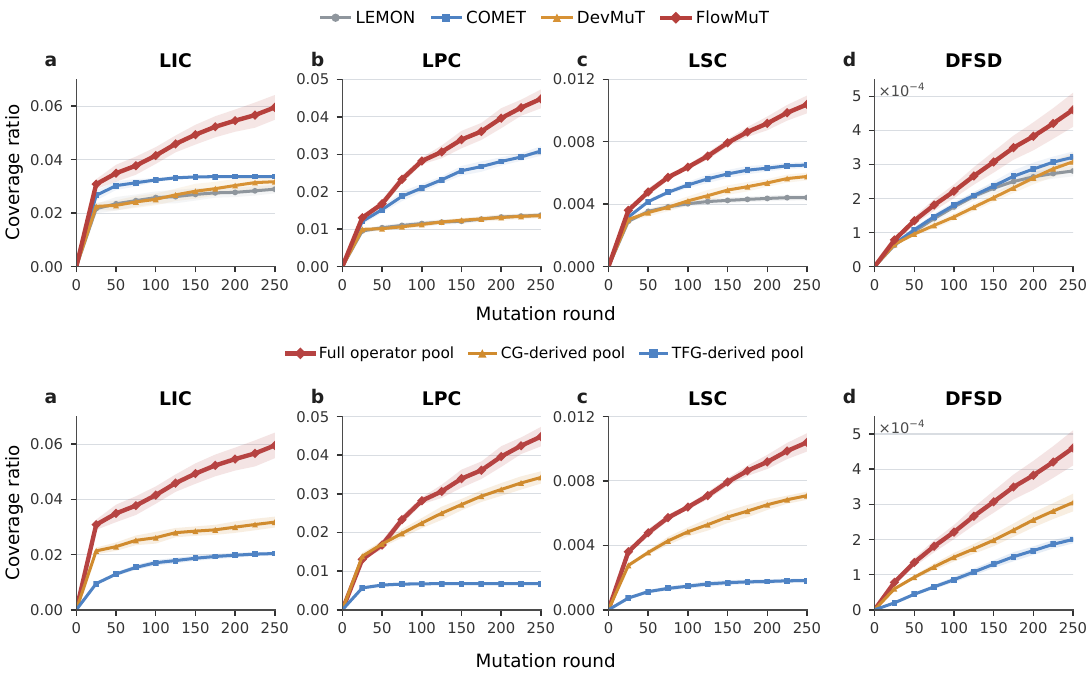}
  \caption{Test case diversity over 250 mutation rounds.  Columns
  (a)--(d) show LIC, LPC, LSC, and DFSD; the rows compare methods (top) and
  operator pools (bottom).  Lines are means with 95.0\% intervals.}
  \Description{Eight line charts form a two-by-four grid.  The columns show
  LIC, LPC, LSC, and DFSD.  The top row compares LEMON, COMET, DevMuT, and
  FlowMuT, while the bottom row compares the full, CG-derived, and TFG-derived
  operator pools.}
  \label{fig:rq2-coverage}
\end{figure}

\parlabel{Comparison with baselines.}
Table~\ref{tab:comparison} reports the final test case diversity, framework
coverage, and cost results, while the top row of Figure~\ref{fig:rq2-coverage}
shows diversity over time.

\begin{table}[!htbp]
\caption{Test case diversity, framework coverage, and execution cost of \tool
and the three baseline methods at mutation round 250.}
\label{tab:comparison}
\centering
\scriptsize
\setlength{\tabcolsep}{3.0pt}
\resizebox{\linewidth}{!}{%
\begin{tabular}{lccccccccc}
\toprule
\rowcolor{tableheader}
& \multicolumn{4}{c}{Test Case Diversity at Round 250} &
\multicolumn{3}{c}{Framework Coverage (\%)} &
\multicolumn{2}{c}{Execution Time (s/test case)} \\
\cmidrule(lr){2-5}\cmidrule(lr){6-8}\cmidrule(lr){9-10}
\rowcolor{tableheader}
Method & \shortstack{LIC\\($\times10^{-2}$)} &
\shortstack{LPC\\($\times10^{-2}$)} &
\shortstack{LSC\\($\times10^{-2}$)} &
\shortstack{DFSD\\($\times10^{-5}$)} &
Python & API & C++ &
Generation & Detection \\
\midrule
LEMON &
2.89 & 1.37 & 0.44 & 28.08 &
27.8 &
28.0 &
34.8 &
\textbf{0.85} &
\textbf{3.19} \\
COMET &
3.37 & 3.09 & 0.65 & 32.11 &
28.0 &
28.1 &
35.0 &
7.60 &
3.76 \\
DevMuT &
3.17 & 1.36 & 0.58 & 30.80 &
30.8 &
32.0 &
41.5 &
1.20 &
3.98 \\
\rowcolor{tablehighlight}
\textbf{\tool} &
\textbf{5.95} & \textbf{4.48} & \textbf{1.04} &
\textbf{46.06} &
\textbf{37.8} &
\textbf{42.2} &
\textbf{48.1} &
6.04 &
4.58 \\
\bottomrule
\end{tabular}
}
\end{table}

\begingroup\looseness=-1
\tool leads every test case diversity and framework-coverage metric.  Its
especially large gains in LPC (45.0\%) and DFSD (43.4\%) indicate that data flow
guidance expands both parameter settings and execution states; the C++
source-line overlap
in the single-run example of Figure~\ref{fig:coverage-bug-overlap}(b) further
shows that these executions reach additional implementation code.  Its
combined generation and detection time is 10.63 seconds per test case, lower
than COMET but higher than LEMON
and DevMuT.\par\endgroup

\parlabel{Mutation operator contributions.}
Table~\ref{tab:operator-pipeline-analysis} and
the bottom row of Figure~\ref{fig:rq2-coverage} report the operator-pool
ablations and their diversity trajectories.  The full pool achieves the
highest test case diversity and framework coverage across the reported
configurations.
The CG-derived pool exceeds the TFG-derived pool on every reported metric.
Within the TFG-derived pool, removing any listed object type or mutation
primitive lowers LIC, DFSD, and all three framework-coverage measures, while
some reduced pools improve LPC or LSC.

\begin{table}[!t]
  \caption{Test case diversity and framework coverage results for operator-pool
  and TFG ablations at mutation round 250.}
  \label{tab:operator-pipeline-analysis}
  \centering
  \footnotesize
  \setlength{\tabcolsep}{1.5pt}
  \renewcommand{\arraystretch}{1.03}
  \begin{tabular}{P{0.27\linewidth}r*{6}{r}R{0.07\linewidth}}
    \toprule
    \rowcolor{tableheader}
    Configuration & Ops &
    \multicolumn{4}{c}{Test Case Diversity} &
    \multicolumn{3}{c}{Framework Coverage (\%)} \\
    \rowcolor{tableheader}
    Configuration & & \shortstack{LIC\\($\times10^{-2}$)} &
    \shortstack{LPC\\($\times10^{-2}$)} &
    \shortstack{LSC\\($\times10^{-2}$)} &
    \shortstack{DFSD\\($\times10^{-5}$)} & Python & API & C++ \\
    \midrule
    \rowcolor{tablesection}
    \multicolumn{9}{l}{\emph{Operator-source ablation}} \\
    \rowcolor{tableunion!70}
    \textbf{Full operator pool} & 147 & 5.95 & 4.48 &
    1.04 & \textbf{46.06} & 37.8 & 42.2 & 48.1 \\
    \rowcolor{tablecg}
    CG-derived pool & 120 & 3.17 & 3.42 &
    0.71 & 30.51 & 29.9 & 32.4 & 37.7 \\
    \rowcolor{tabletfg}
    TFG-derived pool & 48 & 2.04 & 0.68 &
    0.18 & 20.01 & 20.1 & 20.1 & 21.8 \\
    \midrule
    \rowcolor{tablesection}
    \multicolumn{9}{l}{\emph{TFG object ablation (within the TFG-derived pool)}} \\
    \rowcolor{tabletfg}
    TFG-derived pool & 48 & 2.04 & 0.68 &
    0.18 & 20.01 & 20.1 & 20.1 & 21.8 \\
    w/o Tensor operators & 36 & 1.94 & 0.64 &
    0.21 & 13.32 & 13.4 & 11.3 & 11.6 \\
    w/o Operator operators & 38 & 1.92 & 1.60 &
    0.19 & 11.72 & 7.6 & 6.8 & 8.6 \\
    w/o Interface operators & 35 & 1.76 & 1.89 &
    0.37 & 19.15 & 7.5 & 18.4 & 15.5 \\
    w/o Bounded-subgraph operators & 35 & 1.00 & 0.44 &
    0.17 & 9.04 & 8.0 & 9.3 & 11.0 \\
    \midrule
    \rowcolor{tablesection}
    \multicolumn{9}{l}{\emph{Mutation-primitive ablation (within the TFG-derived pool)}} \\
    w/o Update & 28 & 0.82 & 0.20 &
    0.26 & 7.26 & 15.7 & 9.8 & 15.8 \\
    w/o Insertion & 32 & 1.50 & 1.13 &
    0.19 & 18.07 & 13.6 & 17.0 & 10.9 \\
    w/o Deletion & 43 & 1.36 & 1.56 &
    0.19 & 10.01 & 10.4 & 7.7 & 8.0 \\
    w/o Rewiring & 41 & 1.85 & 0.60 &
    0.13 & 19.92 & 12.7 & 15.7 & 15.4 \\
    \bottomrule
  \end{tabular}%
\end{table}

\rqanswer{RQ2}{\tool achieves the highest test case diversity and framework
coverage, including 43.4\% higher DFSD than the strongest baseline.  The full
pool achieves the highest test case diversity and framework coverage among the
operator-pool configurations.}

\subsection{RQ3: Mutation Legality}
\label{sec:rq3}

We examine how each constraint type affects mutation legality and filtering
accuracy.  LMR measures the proportion of legal retained candidates, while
filter selectivity measures how reliably legal and illegal candidates are
distinguished.

\begin{table}[!htbp]
  \caption{Mutation legality and filter selectivity under constraint
  ablations.  TN/FN count legal/illegal retained candidates; FP/TP count
  legal/illegal rejected candidates.}
  \label{tab:constraint-analysis}
  \centering
  \footnotesize
  \setlength{\tabcolsep}{4pt}
  \renewcommand{\arraystretch}{1.08}
  \begin{tabular}{P{0.22\linewidth}*{10}{r}}
    \toprule
    \rowcolor{tableheader}
    & \multicolumn{2}{c}{Mutation legality} &
    \multicolumn{4}{c}{Filter outcomes} &
    \multicolumn{4}{c}{Filter selectivity (\%)} \\
    \cmidrule(lr){2-3}\cmidrule(lr){4-7}\cmidrule(lr){8-11}
    \rowcolor{tableheader}
    Configuration & Ret. (\%) & LMR (\%) & TN & FN & FP & TP &
    Prec. & Rec. & FRR & Leak. \\
    \midrule
    \rowcolor{tablehighlight}
    Full constraints & 88.0 & 96.7 & 851 & 29 & 13 & 107 & 89.2 & 78.7 & 1.5 & 21.3 \\
    w/o input & 92.7 & 92.3 & 856 & 71 & 8 & 65 & 89.0 & 47.8 & 0.9 & 52.2 \\
    w/o internal & 90.7 & 94.2 & 854 & 53 & 10 & 83 & 89.2 & 61.0 & 1.2 & 39.0 \\
    w/o output & 91.6 & 93.3 & 855 & 61 & 9 & 75 & 89.3 & 55.1 & 1.0 & 44.9 \\
    w/o all & 100.0 & 86.4 & 864 & 136 & 0 & 0 & \textit{N/A} & 0.0 & 0.0 & 100.0 \\
    \bottomrule
  \end{tabular}
\end{table}

\begingroup\looseness=-1
\parlabel{Selectivity and constraint contributions.}
Table~\ref{tab:constraint-analysis} shows that the complete system provides the
best selectivity tradeoff, with 89.2\% rejection precision, 78.7\% recall, and
a 1.5\% false-rejection rate.  Removing any constraint type lowers legality
and recall and increases leakage.  Removing input constraints causes the
largest leakage, showing that the three types complement one another.\par\endgroup

\rqanswer{RQ3}{All three constraint types contribute.  Together, they raise the
legal-mutant rate from 86.4\% to 96.7\% while retaining 88.0\% of candidates.}

\subsection{RQ4: Mutation Selection}

We compare selection policies and examine LinUCB's context features using test
case diversity and framework coverage.  Distinct signature counts additionally
show how many different data flow states the generated tests exercise.

\begin{table}[!htbp]
  \caption{Test case diversity and framework coverage under selection-policy
  and LinUCB context-feature ablations at mutation round 250.}
  \label{tab:selection-analysis}
  \centering
  \footnotesize
  \setlength{\tabcolsep}{3pt}
  \renewcommand{\arraystretch}{0.88}
  \begin{tabular}{P{0.20\linewidth}*{3}{C{0.09\linewidth}}C{0.10\linewidth}*{3}{C{0.09\linewidth}}}
    \toprule
    \rowcolor{tableheader}
    & \multicolumn{4}{c}{Test Case Diversity} &
    \multicolumn{2}{c}{Framework Coverage (\%)} &
    \shortstack{Distinct\\signatures} \\
    \cmidrule(lr){2-5}\cmidrule(lr){6-7}
    \rowcolor{tableheader}
    Configuration & \shortstack{LIC\\($\times10^{-2}$)} &
    \shortstack{LPC\\($\times10^{-2}$)} &
    \shortstack{LSC\\($\times10^{-2}$)} &
    \shortstack{DFSD\\($\times10^{-5}$)} & C++ & API & \\
    \midrule
    \rowcolor{tablesection}
    \multicolumn{8}{l}{\emph{Selection-policy comparison}} \\
    One-shot & 3.68 & 1.32 & 0.41 & 43.63 & 35.5 & 27.3 & 460.9 \\
    Random search & 4.56 & 1.50 & 0.62 & 45.55 & 33.3 & 22.7 & 481.2 \\
    Fixed priority & 1.99 & 0.56 & 0.17 & 10.97 & 28.4 & 17.2 & 115.9 \\
    UCB1 & 4.38 & 1.44 & 0.52 & 41.65 & 32.2 & 20.3 & 440.0 \\
    \rowcolor{tablehighlight}
    \textbf{LinUCB} & \textbf{5.95} & \textbf{4.48} & \textbf{1.04} &
    \textbf{46.06} & \textbf{48.1} & \textbf{42.2} & \textbf{486.5} \\
    \midrule
    \rowcolor{tablesection}
    \multicolumn{8}{l}{\emph{LinUCB context-feature ablation}} \\
    w/o site & 5.88 & 4.22 & 0.90 & 41.63 & 45.4 & 36.7 & 439.7 \\
    w/o tensor & 5.22 & 3.91 & 0.81 & 42.47 & 37.7 & 31.3 & 448.6 \\
    w/o operator & 4.11 & 1.05 & 0.75 & 42.80 & 38.6 & 32.7 & 452.1 \\
    w/o downstream & 4.48 & 4.25 & 0.97 & 33.39 & 46.5 & 39.8 & 352.7 \\
    w/o influence & 5.79 & 3.81 & 1.01 & 44.05 & 42.1 & 35.2 & 465.3 \\
    Coverage-only & 3.94 & 2.76 & 0.46 & 32.17 & 36.6 & 29.7 & 339.8 \\
    Impact-only & 2.46 & 0.72 & 0.20 & 8.13 & 32.1 & 22.8 & 85.9 \\
    \bottomrule
  \end{tabular}
\end{table}

\parlabel{Comparison with alternative policies.}
Table~\ref{tab:selection-analysis} shows that LinUCB achieves the highest
test case diversity, framework coverage, and mean distinct signature count.
Compared with context-free UCB1, it increases DFSD and the mean signature count
by 10.6\%.  This result supports using candidate context to transfer feedback
as the TFG and candidate set change.  Random search comes closer in DFSD and
signature count, with LinUCB improving both by only 1.1\%.  However, LinUCB
raises C++ line coverage from 33.3\% to 48.1\% and API coverage from 22.7\% to
42.2\%.  Thus, similar numbers of distinct data flow signatures can correspond
to different framework coverage.

\parlabel{Contribution of context features.}
Removing operator coverage causes the largest LPC and LSC losses among the
single-feature ablations, reducing them by 76.6\% and 27.9\%, respectively.
Removing tensor state coverage causes the largest losses in framework
coverage, by 10.4 percentage points for C++ and 10.9 for API coverage.  These
results suggest that operator and tensor state coverage contribute to
different aspects of selection.  Site coverage also contributes to data flow
diversity; its removal lowers DFSD by 9.6\%, although LIC changes little.

\begingroup\looseness=-1
Among the impact features, removing downstream
impact reduces DFSD and the mean signature count by 27.5\%, the largest
declines among the single-feature ablations.  This supports using downstream
reachability in selection.  Removing gradient influence
has a smaller effect on DFSD, reducing it by 4.4\%, but also lowers C++ and API
coverage by 6.0 and 7.0 percentage points.\par\endgroup

The coverage-only variant exceeds impact-only across all reported metrics,
but both trail the complete context.  Together, the two feature groups account
for both past coverage and potential effects on model execution.

\rqanswer{RQ4}{\tool leads in diversity and coverage, with 48.1\% C++ line
coverage and a mean of 486.5 distinct signatures.  Both coverage and impact
features contribute.}

\FloatBarrier
\enlargethispage{\baselineskip}
\raggedbottom
\section{Threats to Validity}

\begingroup\looseness=-1
\parlabel{Internal Validity.} Implementation differences and stochastic
execution may bias the comparison.  We mitigate this threat through
controlled configurations and repeated runs across all methods.  Component
interactions may also
affect ablation results, so we use matched variants and interpret each
component through its primary metric.  \textbf{Construct Validity.} Data flow
signatures may not represent source coverage, and different failure symptoms
may differ across executions.  We mitigate this threat by reporting diversity,
coverage, and detected bugs separately, and by recording developer confirmation
and fix status for reported bugs.  \textbf{External Validity.} Our evaluation
covers two frameworks and diverse application tasks, but cannot represent
every architecture or data flow pattern.  Future work will evaluate additional
frameworks, accelerators, and deployment environments.\par\endgroup

\section{Conclusion}

We propose \tool, a model-level DL framework testing method grounded in data
flow.  Its \tfg records operator topology, runtime tensor values and properties,
and operator input requirements.  This shared description supports mutation
construction, legality checking, and selection with LinUCB using coverage and
impact features.  Execution feedback guides subsequent selection, and each
accepted mutant provides the TFG for the next iteration.  On PyTorch and
MindSpore, \tool submitted 55 bug reports for cases missed by the baselines,
improved data flow diversity by 43.4\% over the strongest baseline, and
achieved a legal-mutant rate of 96.7\%.

\section{Data Availability}

The implementation and all experimental results, including configurations,
mutants, triggering test cases, and bug reports, are available in the
accompanying artifact~\cite{flowmut2026artifact}.  We also encourage researchers
to contribute new mutation rules following this study.

\bibliographystyle{ACM-Reference-Format}
\bibliography{references}

@inproceedings{abadi2016tensorflow,
  author    = {Mart{\'\i}n Abadi and Paul Barham and Jianmin Chen and Zhifeng Chen and Andy Davis and Jeffrey Dean and Matthieu Devin and Sanjay Ghemawat and Geoffrey Irving and Michael Isard and Manjunath Kudlur and Josh Levenberg and Rajat Monga and Sherry Moore and Derek G. Murray and Benoit Steiner and Paul Tucker and Vijay Vasudevan and Pete Warden and Martin Wicke and Yuan Yu and Xiaoqiang Zheng},
  title     = {{TensorFlow}: A System for Large-Scale Machine Learning},
  booktitle = {Proceedings of the 12th USENIX Symposium on Operating Systems Design and Implementation},
  year      = {2016},
  pages     = {265--283},
  publisher = {USENIX Association},
  url       = {https://www.usenix.org/conference/osdi16/technical-sessions/presentation/abadi}
}

@inproceedings{ansel2024pytorch2,
  author    = {Jason Ansel and Edward Yang and Horace He and Natalia Gimelshein and Animesh Jain and Michael Voznesensky and Bin Bao and Peter Bell and David Berard and Evgeni Burovski and Geeta Chauhan and Anjali Chourdia and Will Constable and Alban Desmaison and Zachary DeVito and Elias Ellison and Will Feng and Jiong Gong and Michael Gschwind and Brian Hirsh and Sherlock Huang and Kshiteej Kalambarkar and Laurent Kirsch and Michael Lazos and Mario Lezcano and Yanbo Liang and Jason Liang and Yinghai Lu and CK Luk and Bert Maher and Yunjie Pan and Christian Puhrsch and Matthias Reso and Mark Saroufim and Marcos Yukio Siraichi and Helen Suk and Michael Suo and Phil Tillet and Eikan Wang and Xiaodong Wang and William Wen and Shunting Zhang and Xu Zhao and Keren Zhou and Richard Zou and Ajit Mathews and Gregory Chanan and Peng Wu and Soumith Chintala},
  title     = {{PyTorch 2}: Faster Machine Learning Through Dynamic Python Bytecode Transformation and Graph Compilation},
  booktitle = {Proceedings of the 29th ACM International Conference on Architectural Support for Programming Languages and Operating Systems, Volume 2},
  year      = {2024},
  pages     = {929--947},
  publisher = {Association for Computing Machinery},
  doi       = {10.1145/3620665.3640366}
}

@inproceedings{li2010contextual,
  author    = {Lihong Li and Wei Chu and John Langford and Robert E. Schapire},
  title     = {A Contextual-Bandit Approach to Personalized News Article Recommendation},
  booktitle = {Proceedings of the 19th International Conference on World Wide Web},
  year      = {2010},
  pages     = {661--670},
  publisher = {Association for Computing Machinery},
  doi       = {10.1145/1772690.1772758}
}

@inproceedings{paszke2019pytorch,
  author    = {Adam Paszke and Sam Gross and Francisco Massa and Adam Lerer and James Bradbury and Gregory Chanan and Trevor Killeen and Zeming Lin and Natalia Gimelshein and Luca Antiga and Alban Desmaison and Andreas Kopf and Edward Yang and Zachary DeVito and Martin Raison and Alykhan Tejani and Sasank Chilamkurthy and Benoit Steiner and Lu Fang and Junjie Bai and Soumith Chintala},
  title     = {{PyTorch}: An Imperative Style, High-Performance Deep Learning Library},
  booktitle = {Advances in Neural Information Processing Systems},
  year      = {2019},
  volume    = {32},
  pages     = {8026--8037},
  publisher = {Curran Associates, Inc.},
  url       = {https://proceedings.neurips.cc/paper_files/paper/2019/file/bdbca288fee7f92f2bfa9f7012727740-Paper.pdf}
}

@inproceedings{pham2019cradle,
  author    = {Hung Viet Pham and Thibaud Lutellier and Weizhen Qi and Lin Tan},
  title     = {{CRADLE}: Cross-Backend Validation to Detect and Localize Bugs in Deep Learning Libraries},
  booktitle = {Proceedings of the 41st IEEE/ACM International Conference on Software Engineering},
  year      = {2019},
  pages     = {1027--1038},
  publisher = {Institute of Electrical and Electronics Engineers},
  doi       = {10.1109/ICSE.2019.00107}
}

@inproceedings{wang2022eagle,
  author    = {Jiannan Wang and Thibaud Lutellier and Shangshu Qian and Hung Viet Pham and Lin Tan},
  title     = {{EAGLE}: Creating Equivalent Graphs to Test Deep Learning Libraries},
  booktitle = {Proceedings of the 44th IEEE/ACM International Conference on Software Engineering},
  year      = {2022},
  pages     = {798--810},
  publisher = {Association for Computing Machinery},
  doi       = {10.1145/3510003.3510165}
}

@article{li2026dllens,
  author    = {Meiziniu Li and Dongze Li and Jianmeng Liu and Jialun Cao and Yongqiang Tian and Shing-Chi Cheung},
  title     = {Enhancing Differential Testing with {LLMs} for Testing Deep Learning Libraries},
  journal   = {ACM Transactions on Software Engineering and Methodology},
  year      = {2026},
  volume    = {35},
  number    = {4},
  articleno = {88},
  numpages  = {39},
  doi       = {10.1145/3735637}
}

@inproceedings{wang2020lemon,
  author    = {Zan Wang and Ming Yan and Junjie Chen and Shuang Liu and Dongdi Zhang},
  title     = {Deep Learning Library Testing via Effective Model Generation},
  booktitle = {Proceedings of the 28th ACM Joint Meeting on European Software Engineering Conference and Symposium on the Foundations of Software Engineering},
  year      = {2020},
  pages     = {788--799},
  publisher = {Association for Computing Machinery},
  doi       = {10.1145/3368089.3409761}
}

@inproceedings{guo2020audee,
  author    = {Qianyu Guo and Xiaofei Xie and Yi Li and Xiaoyu Zhang and Yang Liu and Xiaohong Li and Chao Shen},
  title     = {{Audee}: Automated Testing for Deep Learning Frameworks},
  booktitle = {Proceedings of the 35th IEEE/ACM International Conference on Automated Software Engineering},
  year      = {2020},
  pages     = {486--498},
  publisher = {Association for Computing Machinery},
  doi       = {10.1145/3324884.3416571}
}

@inproceedings{gu2022muffin,
  author    = {Jiazhen Gu and Xuchuan Luo and Yangfan Zhou and Xin Wang},
  title     = {{Muffin}: Testing Deep Learning Libraries via Neural Architecture Fuzzing},
  booktitle = {Proceedings of the 44th IEEE/ACM International Conference on Software Engineering},
  year      = {2022},
  pages     = {1418--1430},
  publisher = {Association for Computing Machinery},
  doi       = {10.1145/3510003.3510092}
}

@article{li2023comet,
  author    = {Meiziniu Li and Jialun Cao and Yongqiang Tian and Tsz On Li and Ming Wen and Shing-Chi Cheung},
  title     = {{COMET}: Coverage-Guided Model Generation for Deep Learning Library Testing},
  journal   = {ACM Transactions on Software Engineering and Methodology},
  year      = {2023},
  volume    = {32},
  number    = {5},
  articleno = {127},
  numpages  = {34},
  doi       = {10.1145/3583566}
}

@inproceedings{wei2022freefuzz,
  author    = {Anjiang Wei and Yinlin Deng and Chenyuan Yang and Lingming Zhang},
  title     = {Free Lunch for Testing: Fuzzing Deep-Learning Libraries from Open Source},
  booktitle = {Proceedings of the 44th IEEE/ACM International Conference on Software Engineering},
  year      = {2022},
  pages     = {995--1007},
  publisher = {Association for Computing Machinery},
  doi       = {10.1145/3510003.3510041}
}

@article{harzevili2025orion,
  author    = {Nima Shiri Harzevili and Mohammad Mahdi Mohajer and Moshi Wei and Hung Viet Pham and Song Wang},
  title     = {History-Driven Fuzzing for Deep Learning Libraries},
  journal   = {ACM Transactions on Software Engineering and Methodology},
  year      = {2025},
  volume    = {34},
  number    = {1},
  articleno = {19},
  numpages  = {29},
  doi       = {10.1145/3688838}
}

@inproceedings{liu2023nnsmith,
  author    = {Jiawei Liu and Jinkun Lin and Fabian Ruffy and Cheng Tan and Jinyang Li and Aurojit Panda and Lingming Zhang},
  title     = {{NNSmith}: Generating Diverse and Valid Test Cases for Deep Learning Compilers},
  booktitle = {Proceedings of the 28th ACM International Conference on Architectural Support for Programming Languages and Operating Systems, Volume 2},
  year      = {2023},
  pages     = {530--543},
  publisher = {Association for Computing Machinery},
  doi       = {10.1145/3575693.3575707}
}

@inproceedings{deng2023titanfuzz,
  author    = {Yinlin Deng and Chunqiu Steven Xia and Haoran Peng and Chenyuan Yang and Lingming Zhang},
  title     = {Large Language Models Are Zero-Shot Fuzzers: Fuzzing Deep-Learning Libraries via Large Language Models},
  booktitle = {Proceedings of the 32nd ACM SIGSOFT International Symposium on Software Testing and Analysis},
  year      = {2023},
  pages     = {423--435},
  publisher = {Association for Computing Machinery},
  doi       = {10.1145/3597926.3598067}
}

@inproceedings{naziri2026centaur,
  author    = {M. M. Abid Naziri and Shinhae Kim and Feiran (Alex) Qin and Marcelo d'Amorim and Saikat Dutta},
  title     = {Testing Deep Learning Libraries via Neurosymbolic Constraint Learning},
  booktitle = {Proceedings of the 48th IEEE/ACM International Conference on Software Engineering},
  year      = {2026},
  pages     = {3586--3598},
  publisher = {Association for Computing Machinery},
  doi       = {10.1145/3744916.3787800}
}

@inproceedings{kim2025pathfinder,
  author    = {Sehoon Kim and Yonghyeon Kim and Dahyeon Park and Yuseok Jeon and Jooyong Yi and Mijung Kim},
  title     = {Lightweight Concolic Testing via Path-Condition Synthesis for Deep Learning Libraries},
  booktitle = {Proceedings of the 47th IEEE/ACM International Conference on Software Engineering},
  year      = {2025},
  pages     = {2957--2969},
  publisher = {Institute of Electrical and Electronics Engineers},
  doi       = {10.1109/ICSE55347.2025.00202}
}

@inproceedings{yang2023nablafuzz,
  author    = {Chenyuan Yang and Yinlin Deng and Jiayi Yao and Yuxing Tu and Hanchi Li and Lingming Zhang},
  title     = {Fuzzing Automatic Differentiation in Deep-Learning Libraries},
  booktitle = {Proceedings of the 45th IEEE/ACM International Conference on Software Engineering},
  year      = {2023},
  pages     = {1174--1186},
  publisher = {Institute of Electrical and Electronics Engineers},
  doi       = {10.1109/ICSE48619.2023.00105}
}

@article{zhang2025dllibrarysurvey,
  author    = {Xiaoyu Zhang and Weipeng Jiang and Chao Shen and Qi Li and Qian Wang and Chenhao Lin and Xiaohong Guan},
  title     = {Deep Learning Library Testing: Definition, Methods and Challenges},
  journal   = {ACM Computing Surveys},
  year      = {2025},
  volume    = {57},
  number    = {7},
  articleno = {187},
  numpages  = {37},
  doi       = {10.1145/3716497}
}

@inproceedings{chen2018tvm,
  author    = {Tianqi Chen and Thierry Moreau and Ziheng Jiang and Lianmin Zheng and Eddie Yan and Haichen Shen and Meghan Cowan and Leyuan Wang and Yuwei Hu and Luis Ceze and Carlos Guestrin and Arvind Krishnamurthy},
  title     = {{TVM}: An Automated End-to-End Optimizing Compiler for Deep Learning},
  booktitle = {Proceedings of the 13th USENIX Symposium on Operating Systems Design and Implementation},
  year      = {2018},
  pages     = {578--594},
  publisher = {USENIX Association},
  url       = {https://www.usenix.org/conference/osdi18/presentation/chen}
}

@article{mu2026howfar,
  author    = {Yanzhou Mu and Rong Wang and Juan Zhai and Chunrong Fang and Xiang Chen and Peiran Yang and Zhixiang Cao and Ruixiang Qian and Shaoyu Yang and Zhenyu Chen},
  title     = {Deep Learning Framework Testing via Model Mutation: How Far Are We?},
  journal   = {IEEE Transactions on Software Engineering},
  year      = {2026},
  volume    = {52},
  number    = {3},
  pages     = {835--854},
  doi       = {10.1109/TSE.2026.3651022}
}

@article{xie2026sort,
  author    = {Xiaoyuan Xie and Yan Song and Songqiang Chen and Jinfu Chen},
  title     = {Subgraph-Oriented Testing for Deep Learning Libraries},
  journal   = {IEEE Transactions on Software Engineering},
  year      = {2026},
  volume    = {52},
  number    = {3},
  pages     = {908--922},
  doi       = {10.1109/TSE.2026.3655712}
}

@inproceedings{mu2024devmut,
  author    = {Yanzhou Mu and Juan Zhai and Chunrong Fang and Xiang Chen and Zhixiang Cao and Peiran Yang and Yinglong Zou and Tao Zheng and Zhenyu Chen},
  title     = {{DevMuT}: Testing Deep Learning Framework via Developer Expertise-Based Mutation},
  booktitle = {Proceedings of the 39th IEEE/ACM International Conference on Automated Software Engineering},
  year      = {2024},
  pages     = {1533--1544},
  publisher = {Association for Computing Machinery},
  doi       = {10.1145/3691620.3695523}
}

@inproceedings{woo2023convnextv2,
  author    = {Sanghyun Woo and Shoubhik Debnath and Ronghang Hu and Xinlei Chen and Zhuang Liu and In So Kweon and Saining Xie},
  title     = {{ConvNeXt V2}: Co-designing and Scaling ConvNets with Masked Autoencoders},
  booktitle = {Proceedings of the IEEE/CVF Conference on Computer Vision and Pattern Recognition},
  year      = {2023},
  pages     = {16133--16142},
  publisher = {Institute of Electrical and Electronics Engineers},
  doi       = {10.1109/CVPR52729.2023.01548}
}

@inproceedings{liu2022swinv2,
  author    = {Ze Liu and Han Hu and Yutong Lin and Zhuliang Yao and Zhenda Xie and Yixuan Wei and Jia Ning and Yue Cao and Zheng Zhang and Li Dong and Furu Wei and Baining Guo},
  title     = {Swin Transformer {V2}: Scaling Up Capacity and Resolution},
  booktitle = {Proceedings of the IEEE/CVF Conference on Computer Vision and Pattern Recognition},
  year      = {2022},
  pages     = {11999--12009},
  publisher = {Institute of Electrical and Electronics Engineers},
  doi       = {10.1109/CVPR52688.2022.01170}
}

@inproceedings{tan2021efficientnetv2,
  author    = {Mingxing Tan and Quoc Le},
  title     = {{EfficientNetV2}: Smaller Models and Faster Training},
  booktitle = {Proceedings of the 38th International Conference on Machine Learning},
  year      = {2021},
  series    = {Proceedings of Machine Learning Research},
  volume    = {139},
  pages     = {10096--10106},
  publisher = {PMLR},
  url       = {https://proceedings.mlr.press/v139/tan21a.html}
}

@inproceedings{wang2024yolov10,
  author    = {Ao Wang and Hui Chen and Lihao Liu and Kai Chen and Zijia Lin and Jungong Han and Guiguang Ding},
  title     = {{YOLOv10}: Real-Time End-to-End Object Detection},
  booktitle = {Advances in Neural Information Processing Systems},
  year      = {2024},
  volume    = {37},
  pages     = {107984--108011},
  publisher = {Curran Associates, Inc.},
  doi       = {10.52202/079017-3429}
}

@inproceedings{xie2021segformer,
  author    = {Enze Xie and Wenhai Wang and Zhiding Yu and Anima Anandkumar and Jose M. Alvarez and Ping Luo},
  title     = {{SegFormer}: Simple and Efficient Design for Semantic Segmentation with Transformers},
  booktitle = {Advances in Neural Information Processing Systems},
  year      = {2021},
  volume    = {34},
  pages     = {12077--12090},
  publisher = {Curran Associates, Inc.},
  url       = {https://papers.nips.cc/paper/2021/hash/64f1f27bf1b4ec22924fd0acb550c235-Abstract.html}
}

@inproceedings{cheng2022mask2former,
  author    = {Bowen Cheng and Ishan Misra and Alexander G. Schwing and Alexander Kirillov and Rohit Girdhar},
  title     = {Masked-Attention Mask Transformer for Universal Image Segmentation},
  booktitle = {Proceedings of the IEEE/CVF Conference on Computer Vision and Pattern Recognition},
  year      = {2022},
  pages     = {1280--1289},
  publisher = {Institute of Electrical and Electronics Engineers},
  doi       = {10.1109/CVPR52688.2022.00135}
}

@inproceedings{warner2025modernbert,
  author    = {Benjamin Warner and Antoine Chaffin and Benjamin Clavi{\'e} and Orion Weller and Oskar Hallstr{\"o}m and Said Taghadouini and Alexis Gallagher and Raja Biswas and Faisal Ladhak and Tom Aarsen and Griffin Thomas Adams and Jeremy Howard and Iacopo Poli},
  title     = {Smarter, Better, Faster, Longer: A Modern Bidirectional Encoder for Fast, Memory Efficient, and Long Context Finetuning and Inference},
  booktitle = {Proceedings of the 63rd Annual Meeting of the Association for Computational Linguistics (Volume 1: Long Papers)},
  year      = {2025},
  pages     = {2526--2547},
  publisher = {Association for Computational Linguistics},
  doi       = {10.18653/v1/2025.acl-long.127}
}

@inproceedings{batzner2024efficientad,
  author    = {Kilian Batzner and Lars Heckler and Rebecca K{\"o}nig},
  title     = {{EfficientAD}: Accurate Visual Anomaly Detection at Millisecond-Level Latencies},
  booktitle = {Proceedings of the IEEE/CVF Winter Conference on Applications of Computer Vision},
  year      = {2024},
  pages     = {127--137},
  publisher = {Institute of Electrical and Electronics Engineers},
  doi       = {10.1109/WACV57701.2024.00020}
}

@inproceedings{bautista2022parseq,
  author    = {Darwin Bautista and Rowel Atienza},
  title     = {Scene Text Recognition with Permuted Autoregressive Sequence Models},
  booktitle = {Proceedings of the 17th European Conference on Computer Vision, Part XXVIII},
  series    = {Lecture Notes in Computer Science},
  volume    = {13688},
  year      = {2022},
  pages     = {178--196},
  publisher = {Springer Nature Switzerland},
  doi       = {10.1007/978-3-031-19815-1_11}
}

@inproceedings{peebles2023dit,
  author    = {William Peebles and Saining Xie},
  title     = {Scalable Diffusion Models with Transformers},
  booktitle = {Proceedings of the IEEE/CVF International Conference on Computer Vision},
  year      = {2023},
  pages     = {4172--4182},
  publisher = {Institute of Electrical and Electronics Engineers},
  doi       = {10.1109/ICCV51070.2023.00387}
}

@inproceedings{liang2021swinir,
  author    = {Jingyun Liang and Jiezhang Cao and Guolei Sun and Kai Zhang and Luc Van Gool and Radu Timofte},
  title     = {{SwinIR}: Image Restoration Using Swin Transformer},
  booktitle = {Proceedings of the IEEE/CVF International Conference on Computer Vision Workshops},
  year      = {2021},
  pages     = {1833--1844},
  publisher = {Institute of Electrical and Electronics Engineers},
  doi       = {10.1109/ICCVW54120.2021.00210}
}

@misc{cyphers2018ngraph,
  author  = {Scott Cyphers and Arjun K. Bansal and Anahita Bhiwandiwalla and Jayaram Bobba and Matthew Brookhart and Avijit Chakraborty and Will Constable and Christian Convey and Leona Cook and Omar Kanawi and Robert Kimball and Jason Knight and Nikolay Korovaiko and Varun Kumar and Yixing Lao and Christopher R. Lishka and Jaikrishnan Menon and Jennifer Myers and Sandeep Aswath Narayana and Adam Procter and Tristan J. Webb},
  title   = {Intel {nGraph}: An Intermediate Representation, Compiler, and Executor for Deep Learning},
  howpublished = {arXiv preprint arXiv:1801.08058},
  year    = {2018},
  doi     = {10.48550/arXiv.1801.08058}
}

@misc{rotem2018glow,
  author  = {Nadav Rotem and Jordan Fix and Saleem Abdulrasool and Garret Catron and Summer Deng and Roman Dzhabarov and Nick Gibson and James Hegeman and Meghan Lele and Roman Levenstein and Jack Montgomery and Bert Maher and Satish Nadathur and Jakob Olesen and Jongsoo Park and Artem Rakhov and Misha Smelyanskiy and Man Wang},
  title   = {{Glow}: Graph Lowering Compiler Techniques for Neural Networks},
  howpublished = {arXiv preprint arXiv:1805.00907},
  year    = {2018},
  doi     = {10.48550/arXiv.1805.00907}
}

@inproceedings{lattner2021mlir,
  author    = {Chris Lattner and Mehdi Amini and Uday Bondhugula and Albert Cohen and Andy Davis and Jacques Pienaar and River Riddle and Tatiana Shpeisman and Nicolas Vasilache and Oleksandr Zinenko},
  title     = {{MLIR}: Scaling Compiler Infrastructure for Domain Specific Computation},
  booktitle = {Proceedings of the IEEE/ACM International Symposium on Code Generation and Optimization},
  year      = {2021},
  pages     = {2--14},
  publisher = {Institute of Electrical and Electronics Engineers},
  doi       = {10.1109/CGO51591.2021.9370308}
}

@article{chen2018metamorphic,
  author    = {Tsong Yueh Chen and Fei-Ching Kuo and Huai Liu and Pak-Lok Poon and Dave Towey and T. H. Tse and Zhi Quan Zhou},
  title     = {Metamorphic Testing: A Review of Challenges and Opportunities},
  journal   = {ACM Computing Surveys},
  year      = {2018},
  volume    = {51},
  number    = {1},
  articleno = {4},
  numpages  = {27},
  doi       = {10.1145/3143561}
}

@inproceedings{zou2024dljsfuzzer,
  author    = {Yinglong Zou and Juan Zhai and Chunrong Fang and Jiawei Liu and Tao Zheng and Zhenyu Chen},
  title     = {Mutation-Based Deep Learning Framework Testing Method in {JavaScript} Environment},
  booktitle = {Proceedings of the 39th IEEE/ACM International Conference on Automated Software Engineering},
  year      = {2024},
  pages     = {970--981},
  publisher = {Association for Computing Machinery},
  doi       = {10.1145/3691620.3695478}
}

@article{liu2024gandalf,
  author    = {Jiawei Liu and Yuheng Huang and Zhijie Wang and Lei Ma and Chunrong Fang and Mingzheng Gu and Xufan Zhang and Zhenyu Chen},
  title     = {Generation-Based Differential Fuzzing for Deep Learning Libraries},
  journal   = {ACM Transactions on Software Engineering and Methodology},
  year      = {2024},
  volume    = {33},
  number    = {2},
  articleno = {50},
  numpages  = {28},
  doi       = {10.1145/3628159}
}

@inproceedings{zhang2025orthrus,
  author    = {Tongwei Zhang and Baojian Hua},
  title     = {{ORThrus}: Detecting Deep Learning Compiler Bugs via Optimization Resistance Transformations},
  booktitle = {Proceedings of the 24th IEEE International Conference on Trust, Security and Privacy in Computing and Communications},
  year      = {2025},
  pages     = {2214--2223},
  publisher = {Institute of Electrical and Electronics Engineers},
  doi       = {10.1109/TrustCom66490.2025.00258}
}

@article{wang2025d3,
  author  = {Jiannan Wang and Hung Viet Pham and Qi Li and Lin Tan and Yu Guo and Adnan Aziz and Erik Meijer},
  title   = {{$D^3$}: Differential Testing of Distributed Deep Learning with Model Generation},
  journal = {IEEE Transactions on Software Engineering},
  year    = {2025},
  volume  = {51},
  number  = {1},
  pages   = {38--52},
  doi     = {10.1109/TSE.2024.3461657}
}

@article{ji2025moco,
  author  = {Pin Ji and Yang Feng and Duo Wu and Lingyue Yan and Penglin Chen and Jia Liu and Zhihong Zhao},
  title   = {{MoCo}: Fuzzing Deep Learning Libraries via Assembling Code},
  journal = {IEEE Transactions on Software Engineering},
  year    = {2025},
  volume  = {51},
  number  = {2},
  pages   = {371--388},
  doi     = {10.1109/TSE.2024.3509975}
}

@inproceedings{christou2023ivysyn,
  author    = {Neophytos Christou and Di Jin and Vaggelis Atlidakis and Baishakhi Ray and Vasileios P. Kemerlis},
  title     = {{IvySyn}: Automated Vulnerability Discovery in Deep Learning Frameworks},
  booktitle = {Proceedings of the 32nd USENIX Security Symposium},
  year      = {2023},
  pages     = {2383--2400},
  publisher = {USENIX Association},
  url       = {https://www.usenix.org/conference/usenixsecurity23/presentation/christou}
}

@article{liu2022tzer,
  author    = {Jiawei Liu and Yuxiang Wei and Sen Yang and Yinlin Deng and Lingming Zhang},
  title     = {Coverage-Guided Tensor Compiler Fuzzing with Joint {IR}-Pass Mutation},
  journal   = {Proceedings of the ACM on Programming Languages},
  year      = {2022},
  volume    = {6},
  number    = {OOPSLA1},
  articleno = {73},
  numpages  = {26},
  doi       = {10.1145/3527317}
}

@inproceedings{su2023torchprobe,
  author    = {Qidong Su and Chuqin Geng and Gennady Pekhimenko and Xujie Si},
  title     = {{TorchProbe}: Fuzzing Dynamic Deep Learning Compilers},
  booktitle = {Proceedings of the 21st Asian Symposium on Programming Languages and Systems},
  series    = {Lecture Notes in Computer Science},
  volume    = {14405},
  year      = {2023},
  pages     = {310--331},
  publisher = {Springer Nature Singapore},
  doi       = {10.1007/978-981-99-8311-7_15}
}

@inproceedings{ma2023hirgen,
  author    = {Haoyang Ma and Qingchao Shen and Yongqiang Tian and Junjie Chen and Shing-Chi Cheung},
  title     = {Fuzzing Deep Learning Compilers with {HirGen}},
  booktitle = {Proceedings of the 32nd ACM SIGSOFT International Symposium on Software Testing and Analysis},
  year      = {2023},
  pages     = {248--260},
  publisher = {Association for Computing Machinery},
  doi       = {10.1145/3597926.3598053}
}

@inproceedings{liu2023neuri,
  author    = {Jiawei Liu and Jinjun Peng and Yuyao Wang and Lingming Zhang},
  title     = {{NeuRI}: Diversifying {DNN} Generation via Inductive Rule Inference},
  booktitle = {Proceedings of the 31st ACM Joint European Software Engineering Conference and Symposium on the Foundations of Software Engineering},
  year      = {2023},
  pages     = {657--669},
  publisher = {Association for Computing Machinery},
  doi       = {10.1145/3611643.3616337}
}

@misc{lv2024rtdetrv2,
  author  = {Wenyu Lv and Yian Zhao and Qinyao Chang and Kui Huang and Guanzhong Wang and Yi Liu},
  title   = {{RT-DETRv2}: Improved Baseline with Bag-of-Freebies for Real-Time Detection Transformer},
  howpublished = {arXiv preprint arXiv:2407.17140},
  year    = {2024},
  doi     = {10.48550/arXiv.2407.17140}
}

@misc{jiang2023rtmpose,
  author  = {Tao Jiang and Peng Lu and Li Zhang and Ningsheng Ma and Rui Han and Chengqi Lyu and Yining Li and Kai Chen},
  title   = {{RTMPose}: Real-Time Multi-Person Pose Estimation Based on {MMPose}},
  howpublished = {arXiv preprint arXiv:2303.07399},
  year    = {2023},
  doi     = {10.48550/arXiv.2303.07399}
}

@misc{qwen2024qwen25,
  author  = {{Qwen} and An Yang and Baosong Yang and Beichen Zhang and Binyuan Hui and Bo Zheng and Bowen Yu and Chengyuan Li and Dayiheng Liu and Fei Huang and Haoran Wei and Huan Lin and Jian Yang and Jianhong Tu and Jianwei Zhang and Jianxin Yang and Jiaxi Yang and Jingren Zhou and Junyang Lin and Kai Dang and Keming Lu and Keqin Bao and Kexin Yang and Le Yu and Mei Li and Mingfeng Xue and Pei Zhang and Qin Zhu and Rui Men and Runji Lin and Tianhao Li and Tianyi Tang and Tingyu Xia and Xingzhang Ren and Xuancheng Ren and Yang Fan and Yang Su and Yichang Zhang and Yu Wan and Yuqiong Liu and Zeyu Cui and Zhenru Zhang and Zihan Qiu},
  title   = {{Qwen2.5} Technical Report},
  howpublished = {arXiv preprint arXiv:2412.15115},
  year    = {2024},
  doi     = {10.48550/arXiv.2412.15115}
}

@inproceedings{xie2022docter,
  author    = {Danning Xie and Yitong Li and Mijung Kim and Hung Viet Pham and Lin Tan and Xiangyu Zhang and Michael W. Godfrey},
  title     = {{DocTer}: Documentation-Guided Fuzzing for Testing Deep Learning {API} Functions},
  booktitle = {Proceedings of the 31st ACM SIGSOFT International Symposium on Software Testing and Analysis},
  year      = {2022},
  pages     = {176--188},
  publisher = {Association for Computing Machinery},
  doi       = {10.1145/3533767.3534220}
}

@article{zhang2022mltesting,
  author  = {Jie M. Zhang and Mark Harman and Lei Ma and Yang Liu},
  title   = {Machine Learning Testing: Survey, Landscapes and Horizons},
  journal = {IEEE Transactions on Software Engineering},
  year    = {2022},
  volume  = {48},
  number  = {1},
  pages   = {1--36},
  doi     = {10.1109/TSE.2019.2962027}
}

@inproceedings{sun2020transrepair,
  author    = {Zeyu Sun and Jie M. Zhang and Mark Harman and Mike Papadakis and Lu Zhang},
  title     = {Automatic Testing and Improvement of Machine Translation},
  booktitle = {Proceedings of the 42nd IEEE/ACM International Conference on Software Engineering},
  year      = {2020},
  pages     = {974--985},
  publisher = {Association for Computing Machinery},
  doi       = {10.1145/3377811.3380420}
}

@article{shi2025multigranularity,
  author  = {Ying Shi and Zheng Zheng and Beibei Yin and Zhiyu Xi},
  title   = {Multigranularity Coverage Criteria for Deep Learning Libraries},
  journal = {Software Testing, Verification and Reliability},
  year    = {2025},
  volume  = {35},
  number  = {1},
  pages   = {e1903},
  doi     = {10.1002/stvr.1903}
}

@inproceedings{zhou2020metamorphic,
  author    = {Jinyi Zhou and Kun Qiu and Zheng Zheng and Tsong Yueh Chen and Pak-Lok Poon},
  title     = {Using Metamorphic Testing to Evaluate {DNN} Coverage Criteria},
  booktitle = {Proceedings of the 31st IEEE International Symposium on Software Reliability Engineering Workshops},
  year      = {2020},
  pages     = {147--148},
  publisher = {Institute of Electrical and Electronics Engineers},
  doi       = {10.1109/ISSREW51248.2020.00055}
}

@misc{flowmut2026artifact,
  author       = {{Anonymous Authors}},
  title        = {{FlowMuT} Artifact},
  year         = {2026},
  howpublished = {Online repository},
  url          = {https://anonymous.4open.science/r/FlowMuT-2140}
}

@article{zou2023ramos,
  author  = {Yinglong Zou and Haofeng Sun and Chunrong Fang and Jiawei Liu and Zhenping Zhang},
  title   = {Deep Learning Framework Testing via Hierarchical and Heuristic Model Generation},
  journal = {Journal of Systems and Software},
  year    = {2023},
  volume  = {201},
  pages   = {111681},
  doi     = {10.1016/j.jss.2023.111681}
}

@article{mu2025modelmeta,
  author  = {Yanzhou Mu and Juan Zhai and Chunrong Fang and Xiang Chen and Zhixiang Cao and Peiran Yang and Kexin Zhao and An Guo and Zhenyu Chen},
  title   = {Improving Deep Learning Framework Testing with Model-Level Metamorphic Testing},
  journal = {Proceedings of the ACM on Software Engineering},
  year    = {2025},
  volume  = {2},
  number  = {ISSTA},
  pages   = {2158--2180},
  doi     = {10.1145/3728972}
}

@article{zou2026dlmmm,
  author  = {Yinglong Zou and Juan Zhai and Chunrong Fang and Yanzhou Mu and Jiawei Liu and Zhenyu Chen},
  title   = {Deep Learning Framework Testing via Heuristic Guidance Based on Multiple Model Measurements},
  journal = {ACM Transactions on Software Engineering and Methodology},
  year    = {2026},
  note    = {Advance online publication},
  doi     = {10.1145/3830465}
}

@inproceedings{zou2026scalpel,
  author    = {Yinglong Zou and Juan Zhai and Chunrong Fang and An Guo and Jiawei Liu and Zhenyu Chen},
  title     = {{Scalpel}: Automotive Deep Learning Framework Testing via Assembling Model Components},
  booktitle = {Proceedings of the 48th IEEE/ACM International Conference on Software Engineering},
  year      = {2026},
  pages     = {3298--3310},
  publisher = {Association for Computing Machinery},
  doi       = {10.1145/3744916.3773242}
}

@inproceedings{mu2026cipihunter,
  author    = {Yanzhou Mu and Shuo Meng and Mijung Kim and Xiang Chen and Chunrong Fang and Zhenyu Chen and Juan Zhai},
  title     = {{CIPIHunter}: Detecting Configuration-Induced Prediction Instability in Deep Learning Frameworks},
  booktitle = {Proceedings of the 41st IEEE/ACM International Conference on Automated Software Engineering},
  year      = {2026},
  note      = {To appear},
  publisher = {Association for Computing Machinery},
  url       = {https://conf.researchr.org/details/ase-2026/ase-2026-research-track/15/CIPIHunter-Detecting-Configuration-Induced-Prediction-Instability-in-Deep-Learning-F}
}

@article{mu2026llmchallenges,
  author  = {Yanzhou Mu and Rong Wang and Juan Zhai and Chunrong Fang and Xiang Chen and Jiacong Wu and An Guo and Jiawei Shen and Bingzhuo Li and Zhenyu Chen},
  title   = {Understanding {LLM}-Centric Challenges for Deep Learning Frameworks: An Empirical Analysis},
  journal = {ACM Transactions on Software Engineering and Methodology},
  year    = {2026},
  note    = {Advance online publication},
  doi     = {10.1145/3821532}
}

@article{mu2022dmos,
  author  = {Yanzhou Mu and Zan Wang and Xiang Chen and Junjie Chen and Jingke Zhao and Jianmin Wang},
  title   = {Deep Learning Test Optimization Method Using Multi-objective Optimization},
  journal = {International Journal of Software and Informatics},
  year    = {2022},
  volume  = {12},
  number  = {4},
  pages   = {403--436},
  doi     = {10.21655/ijsi.1673-7288.00282}
}

@inproceedings{mu2021hars,
  author    = {Yanzhou Mu and Zan Wang and Shuang Liu and Jun Sun and Junjie Chen and Xiang Chen},
  title     = {{HARS}: Heuristic-Enhanced Adaptive Randomized Scheduling for Concurrency Testing},
  booktitle = {Proceedings of the 21st IEEE International Conference on Software Quality, Reliability and Security},
  year      = {2021},
  pages     = {219--230},
  publisher = {Institute of Electrical and Electronics Engineers},
  doi       = {10.1109/QRS54544.2021.00033}
}

\end{document}